\documentclass[aip,twocolumn,amsmath,amssymb,reprint]{revtex4-2}

\usepackage{graphicx}
\usepackage{dcolumn}
\usepackage{bm}
\usepackage[normalem]{ulem}
\usepackage{xcolor}
\usepackage{adjustbox}
\usepackage{enumitem}
\usepackage{caption}
\newcommand{\be}{\begin{equation}}
\newcommand{\ee}{\end{equation}}
\newcommand{\ba}{\begin{eqnarray}}
\newcommand{\ea}{\end{eqnarray}}

\newcommand{\red}[1]{\textcolor{red}{#1}}
\newcommand{\blue}[1]{\textcolor{blue}{#1}}
\newcommand{\green}[1]{\textcolor{green}{#1}}
\newcommand{\dgreen}[1]{\textcolor{dgreen}{#1}}

\newcommand{\cyan}[1]{\textcolor{cyan}{#1}}

\newcommand{\magenta}[1]{\textcolor{magenta}{#1}}

\newcommand\restr[2]{{
  \left.\kern-\nulldelimiterspace
  #1
  \vphantom{\big|}
  \right|_{#2}
  }}
  
\definecolor{blazeorange}{rgb}{1.0, 0.4, 0.0}
\definecolor{seagreen}{rgb}{0.18, 0.55, 0.34}
\definecolor{rufous}{rgb}{0.66, 0.11, 0.03}
\definecolor{royalfuchsia}{rgb}{0.79, 0.17, 0.57}
\definecolor{scarlet}{rgb}{1.0, 0.13, 0.0}
\definecolor{royalpurple}{rgb}{0.47, 0.32, 0.66}
\definecolor{dpurple}{rgb}{0.4863, 0.2784, 0.5216}
\definecolor{darkblue}{rgb}{0, 0, 0.66}
\definecolor{dgreen}{rgb}{0, 0.85, 0}

\usepackage[utf8]{inputenc}
\usepackage[T1]{fontenc}
\usepackage{mathptmx}
\usepackage{etoolbox}
\usepackage{caption}

\makeatletter
\def\@email#1#2{
 \endgroup
 \patchcmd{\titleblock@produce}
  {\frontmatter@RRAPformat}
  {\frontmatter@RRAPformat{\produce@RRAP{*#1\href{mailto:#2}{#2}}}\frontmatter@RRAPformat}
  {}{}
}
\makeatother

\begin{document}

\preprint{AIP/123-QED}

\title[Oblique Relativistic Shell -- Wall Collision]{Oblique Collision of a Relativistic Cold Shell with an Ideal Reflecting Wall}

\author{Jonathan Granot}\email{granot@openu.ac.il.}
\altaffiliation[Also at ]{Department of Physics, The George Washington University, Washington, DC 20052, USA}
\affiliation{Astrophysics Research Center of the Open university (ARCO), The Open University of Israel, P.O Box 808, Ra’anana 4353701, Israel\looseness=-1}
\affiliation{Astroparticule et Cosmologie (APC), CNRS – Université Paris Cité, F-75013 Paris, France\looseness=-1}
\altaffiliation[Also at ]{Department of Physics, The George Washington University, Washington, DC 20052, USA\looseness=-1}
\author
{Prasanta Bera}
\affiliation{Department of Physics, Institute of Science, Banaras
Hindu University, Varanasi-221005, India\looseness=-1}
\affiliation{Astrophysics Research Center of the Open university (ARCO), The Open University of Israel, P.O Box 808, Ra’anana 4353701, Israel\looseness=-1}

\author
{Michael Rabinovich}
\affiliation{Department of Natural Sciences, The Open University of Israel, P.O Box 808, Ra'anana 4353701, Israel\looseness=-1}
\affiliation{Astrophysics Research Center of the Open university (ARCO), The Open University of Israel, P.O Box 808, Ra’anana 4353701, Israel\looseness=-1}

\author
{Paz Beniamini}
\altaffiliation[Also at ]{Department of Physics, The George Washington University, Washington, DC 20052, USA\looseness=-1}
\affiliation{Department of Natural Sciences, The Open University of Israel, P.O Box 808, Ra'anana 4353701, Israel\looseness=-1}
\affiliation{Astrophysics Research Center of the Open university (ARCO), The Open University of Israel, P.O Box 808, Ra’anana 4353701, Israel\looseness=-1}

\date{\today}

\begin{abstract}
Relativistic flows are common in astrophysics and often form shocks when different parts of the flow collide at relativistic relative velocities. Such collisions are often oblique, forming two shocks whose shocked fluids are separated by a contact discontinuity, which is treated here as an ideal reflecting ``wall'' where the flow on either side is modeled separately. The latter is modeled in the lab frame $S$ as a uniform cold planar shell propagating into vacuum at velocity $v_1=\beta_1c$ normal to its vacuum interface, colliding with the wall at an incidence angle $\alpha_1$. The collision point $P$ moves along the wall at a velocity $v_p=v_{1}/\sin\alpha_1$, and a boost along the wall at $v_p$ leads to a steady-state frame $S'$ where this problem is highly simplified. However, a ``super-luminal'' regime exists where $v_p>c\Leftrightarrow\tan\alpha_1<\Gamma_{1}\beta_{1}=(1-\beta_{1}^2)^{-1/2}\beta_1$ and no steady-state frame $S'$ exists. It corresponds to only very small $\alpha_1$ in the Newtonian regime, but nearly all $\alpha_1$ in the relativistic regime. We solve this problem \textit{\textbf{fully analytically}} using integral conservation laws, in the attachmrnt region where point $P$ is attached to the wall. This region of  parameter space is bound at high $\alpha_1$ by the detachment line, which coincides with the sonic line for a cold initial shell. A weak-shock solution exist in all this region, while a strong-shock solution exists only in the sub-luminal attachment region -- between the luminal line and the detachment/sonic line where the two solutions coincide and beyond which point $P$ detaches from the wall and shocked fluid spills into the vacuum.
\end{abstract}

\maketitle

\section{Introduction}
\label{sec:intro}

Relativistic bulk motion is quite common in astrophysics. Prominent examples include pulsar winds and their nebulae (PWN), jets in active galactic nuclei (AGN), X-ray binaries (or micro-quasars), gamma-ray bursts (GRB) and tidal disruption events (TDE), and outflows associated with magnetar giant flares (MGF) or fast radio bursts (FRB). Such relativistic outflows can lead to collisions between different parts of the flow at relativistic relative velocities. This can occur either within the outflow itself, leading to ``internal shocks'', e.g. in  GRB\cite{rees1994unsteady, sari1997cosmological,daigne1998gamma}, AGN\cite{rees1978m87,spada2001internal} or micro-quasars\citep{kaiser2000internal, malzac2014spectral}, or when such an outflow interacts with the surrounding medium, e.g. in GRB afterglows or MGF outflows.

The two colliding parts of the flow may be approximated as uniform planar shells, which are usually assumed to be co-planar, since this significantly simplifies the problem by making it one-dimensional (1D) \citep[e.g.][]{Peer+17,Rahaman+24a}. Such collisions lead to the formation of a pair of shocks, each propagating into one of the two colliding shells, where the two shocked parts of the shells are separated by a contact discontinuity (CD). However, more generally, such shells are not coplanar and instead can collide at some angle. Such oblique collisions (e.g. \cite{shi2020relativistic}, \cite{Sonkusale+2026arXiv260720978A},\cite{Charlet+2022},\cite{graw2022mildly}), which are the motivation for this work, similarly form a pair of shocks each propagating into one of the two colliding shells where the two shocked regions are separated by a CD. For our purposes it is convenient to treat the CD as a perfect reflecting wall, and model the flow on each of its sides separately. This reduces the problem to two coupled problems of a single shell obliquely colliding with a wall, which could later be matched using the conditions at the CD (of equal pressures and normal component of the velocity across the CD). 

Here we solve the basic problem of a single shell colliding with a perfectly reflecting wall. We use the formalism derived in our previous work \citep[][hereafter Paper~1]{Granot-Rabinovich-24}, which relies on solving the integral conservation laws, and adjust it to our current problem. Paper~1 developed this formalism to solve the problem of relativistic shock reflection for regular reflection, where the point $P$ of collision with the reflecting wall, which moves along the wall at a speed $v_p$, remains attached to the wall. Regular reflection occurs at sufficiently small incidence angles, while beyond a critical detachment angle point $P$ detaches from the wall and there is instead Mach reflection (where point $P$ detaches from the wall and becomes a triple point the is connected to the wall by a Mach stem). There are many analogies between relativistic shock reflection and an oblique cold shell-wall collision that we address here, including the existence of sub-luminal ($v_p<c$) and super-luminal ($v_p>c$) regimes, and a critical detachment angle (where in our case beyond that angle some of the shocked high-pressure fluid spills into the vacuum region).

In \S\,\ref{sec:formulation} we formulate the problem of a cold uniform shell obliquely colliding with a perfect wall, and generalize the integral conservation laws from Paper~1 to it. In \S\,\ref{sec:prev-res}
we outline relevant previous results for a one-dimensional shock, which we later use in this work, and specify to the realistic Taub-Matthews equation of state and a cold initial shell, which later allow for a fully analytic solution. In \S\,\ref{sec:solutions} we derive a fully analytic solution to the problem. In \S\,\ref{sec:results} the results of our solution are shown. In \S\,\ref{sec:cons-steady} we analyze this problem in the steady-state frame $S'$, which exists in the sub-luminal regime ($v_p<c$). In \S\,\ref{sec:simulations} we solve this problem numerically using relativistic hydrodynamic simulations, and show that beyond the detachment line, shocked fluid spills into the vacuum region. Finally, our conclusions are discussed in \S\,\ref{sec:dis}.

\begin{table}
\centering
{\small
\resizebox{0.48\textwidth}{!}{
\begin{tabular}{|c|l|}
\hline
notation & stands for \\
\hline
$P$ & intersection point of the two shocks and the wall\\
$v_p$ & velocity of point $P$ along the wall in the lab frame $S$\\
$S$ & the lab frame, where region 0 and the wall are at rest \\
$S'$ & rest frame of point $P$, where the flow is steady\\
 & (exists only for $v_p<c$) \\
$Q_i$, $Q'_i$
& quantity $Q$ in region $i=0,1,2$ measured in 
frame $S$,\,$S'$
\\
$v_i=\beta_i c$, $v'_i$
& fluid velocity of region $i$ in the rest frame $S$,\,$S'$
\\
$\Gamma_i$, $\Gamma'_i$ 
& Lorentz factor (LF) of region $i$ in rest frame $S$,\,$S'$
\\
$u_i=\Gamma_i\beta_i$, $u'_i$
& proper velocity of region $i$ in rest frame $S$,\,$S'$
\\
$v_{s}=\beta_{s}c$ 
& the shock velocity along its normal in frame $S$  \\
$\Gamma_{s}=(1-\beta_{s}^2)^{-1/2}$ &
the corresponding shock Lorentz factor in frame $S$\\
$u_{s}=\Gamma_{s}\beta_{s}$ &
the corresponding shock proper velocity in frame $S$\\
$\Gamma_{12}=(1-\beta_{12}^2)^{-1/2}$ & Lorentz factor of region $1$ relative to region $2$\\
$v_{12}=\beta_{12}c=-v_{21}$ & relative velocity of regions $1$ and $2$\\
$u_{12}=\Gamma_{12}\beta_{12}=-u_{21}$ & relative proper velocity of regions $1$ and $2$\\
$\beta_{s,i}=-\beta_{i,s}$
& the shock velocity along its normal\\
 & measured in the rest frame of region $i$\\
$\Gamma_{s,i}$, $u_{s,i}$ &  the corresponding Lorentz factor and proper velocity\\
$c_{s,i}=\beta_{c_s,i}c$ & sound speed in region $i$ (in the fluid rest frame)\\
$u_{c_s,i}=(\beta_{c_s,i}^{-2}-1)^{-\frac{1}{2}}$ & the corresponding sound proper speed in region $i$\\
$\rho_{i}$ & proper rest-mass density in region $i$\\
$p_i$ & pressure in region $i$ (measured in the fluid rest frame)\\
$e_{{\rm int},i}$ & the proper internal energy density in region $i$ 
\\
$e_i=e_{{\rm int},i}+\rho_i c^2$ & the proper total energy density in region $i$\\
$\hat{\gamma}_i$ & the adiabatic index of the fluid in region $i$\\
$w_i=e_i+p_i$ & the proper enthalpy density of the fluid in region $i$\\
$h_i=w_i/\rho_i c^2$ & the proper enthalpy per unit rest energy in region $i$\\
\hline
\end{tabular}
}
\caption{Notations used in this work. A quantity (such as density, pressure or time) is said to be ``proper" when it is measured in the fluid rest frame.
Proper velocity (or celerity) is the derivative of the observer-measured location with respect to the proper time $\tau$, i.e. $\textbf{\textit{u}}_ic=
d\textbf{\textit{x}}_i/d\tau_i=
(dt/d\tau_i)(d\textbf{\textit{x}}_i/dt)=
\Gamma_i\bm{\beta}_ic$.
}
\label{tab:notations}
}
\end{table}

\section{Formulating the problem: a cold shell colliding with a wall at an angle}
\label{sec:formulation}

\begin{figure}
\centering
\includegraphics[trim={0cm 0cm 0cm 0cm},clip,scale=0.6,width=0.48\textwidth]{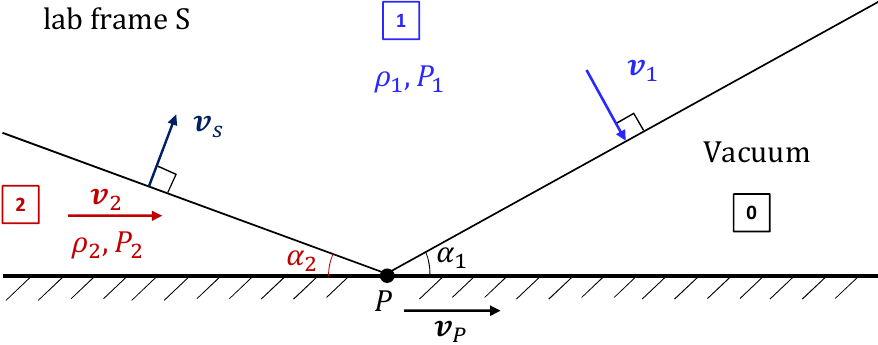}
\\ \vspace{1.0cm}
\includegraphics[trim={0cm 0cm 0cm 0cm},clip,scale=0.6,width=0.48\textwidth]{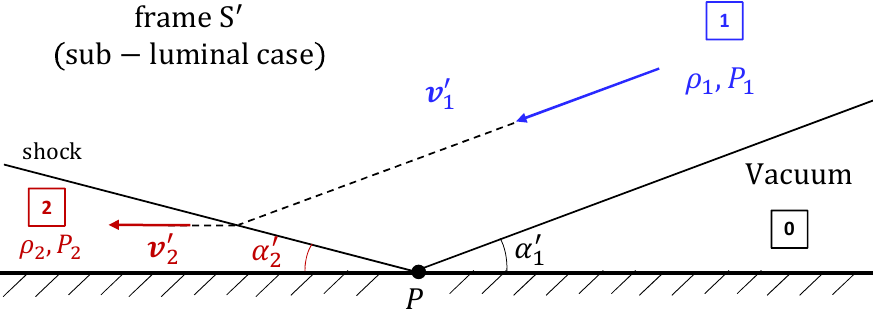}
\vspace{-0.5cm}
\caption{\textit{\textbf{Top}}: In the lab frame $S$ a uniform shell (\textcolor[rgb]{0,0,1}{region 1})  propagates into vacuum (region 0) with a velocity $\textbf{\textit{v}}_1$ normal to its vacuum interface and collides with an ideal reflecting wall at an angle $\alpha_1$. The interaction point $P$ moves along the wall at velocity $\bm{v}_p$, and a shocked \textcolor[rgb]{0.70,0,0}{region 2} is formed moving at a velocity $\bm{v}_2$ along the wall. The shock front is at an angle $\alpha_2$ from the wall and moves at a velocity $\bm{v}_s$ normal to the shock front.
\textit{\textbf{Bottom}}:
In the sub-luminal case ($v_p<c$), one can conveniently transform to the frame $S'$ where point $P$ is at rest and the flow is steady ($v'_p=v'_s=0$) and easier to solve.}
\label{fig:shell-wall}
\end{figure}

The problem considered here is described in Fig.~\ref{fig:shell-wall}. We consider a cold (with pressure\footnote{Strictly speaking, for self-consistency it is required that the shell be cold, $p_1=0$, as otherwise a rarefaction wave would propagate into the shell from the vacuum interface. This is indeed what we assume to derive our result.} $p_1\ll\rho_1c^2u_1^2$ where $u_1=\Gamma_1\beta_1$) uniform (semi-infinite) shell (\textcolor[rgb]{0,0,1}{region 1}) of proper rest mass density $\rho_1$ and velocity $\textbf{\textit{v}}_1=\bm{\beta}_1c$ in the lab frame $S$ that is normal to its front, which is at an angle $\alpha_1$ relative to a rigid wall, and propagates into vacuum
(region 0) and collides with a wall. Due to the collision with the wall, a shock forms that propagates into region 1, whose front in frame $S$ subtends an angle $\alpha_2$ relative to the wall and has a velocity $v_{s}$ (defined along the shock normal). The shocked fluid behind this shock (\textcolor[rgb]{0.7,0,0}{region 2}) moves parallel to the wall at a speed $v_2$, with a pressure $p_2$ and proper rest mass density $\rho_2$. 
The instantaneous collision point $P$
moves along the wall at a velocity
\begin{equation}\label{eq:vp-shell-wall}
v_p = \frac{v_{1}}{\sin\alpha_1} = \frac{v_{s}}{\sin\alpha_2}\ .
\end{equation}

Supersonic fluid motions can lead to shock waves, where at a shock front there is a sharp increase (which may be approximated as a discontinuity) in the density, pressure and temperature of the incoming upstream fluid, such that they are all larger in the shocked downstream region. The specific entropy also increases across the shock, which makes the shock an irreversible process. In the upstream fluid's rest frame (region 1 in our setup), the downstream fluid's velocity just behind the shock ($\bm{v}_{2,1}$) is always along the local shock normal. In this frame, the shock front velocity along its normal, $v_{s,1}$, must be supersonic (exceeding the sound speed $c_{s,1}$ in the upstream region) for a shock to exist. The ratio of the corresponding proper velocity ($u_{s,1}=\Gamma_{s,1}\beta_{s,1}$ where $\Gamma_{s,1}=(1-\beta_{s,1}^2)^{-1/2}$, $\beta_{s,1}=v_{s,1}/c$ and $c$ is the speed of light) to the upstream proper sound speed ($u_{c_s,1}=\Gamma_{c_s,1}\beta_{c_s,1}$ where $\beta_{c_s,1}=c_{s,1}/c$ and $\Gamma_{c_s,1}=(1-\beta_{c_s,1}^2)^{-1/2}$) is defined as the shock's Mach number in this frame, $\mathcal{M}=u_{s,1}/u_{c_s,1}$, and $\mathcal{M}>1$ is required for a shock to exist. For a cold upstream medium that we assume in this work, its sound speed vanishes ($c_{s,1}\to0$) and the shock Mach number becomes infinite ($\mathcal{M}\to\infty$), i.e. we are in the extreme strong shock limit.

Using integral conservation laws, following Paper~1, the equations for shock reflection can be generalized  (upon substitution of $\beta_s$ instead of $\beta_{s2}$). In particular, for the fixed fluid (or rest mass) method, the rate at which volume containing fluid of region 1 (2) decreases (increases) per unit area on the ``wall" is given by
\begin{eqnarray}\nonumber
\dot{\mathbb{V}}_1 &=& -\frac{v_s+v_1\cos\alpha_+}{\cos\alpha_2} = -\frac{v_1}{\cos\alpha_2}\left(\frac{\sin\alpha_2}{\sin\alpha_1}+\cos\alpha_+\right)\ ,\quad\\
\label{eq:fixedM_Vdot}
\dot{\mathbb{V}}_2 &=& \left(\frac{v_s}{\sin\alpha_2}-v_2\right)\tan\alpha_2 = \left(\frac{v_1}{\sin\alpha_1}-v_2\right)\tan\alpha_2\ ,
\end{eqnarray}
where $\alpha_+\equiv\alpha_1+\alpha_2$.
The corresponding equations for conservation of rest mass, energy, and $x$ and $y$ components of the momentum are
\begin{equation}\label{eq:CS-M2}
\rho_1\Gamma_1\dot{\mathbb{V}}_1 + \rho_2\Gamma_2\dot{\mathbb{V}}_2 = 0\ \Leftrightarrow\ 
\frac{\Gamma_2\rho_2}{\Gamma_1\rho_1} = \frac{-\dot{\mathbb{V}}_1}{\dot{\mathbb{V}}_2}=\frac{1+\frac{\sin\alpha_1}{\sin\alpha_2}\cos\alpha_+}{1-\frac{\beta_2}{\beta_1}\sin\alpha_1}\ ,
\end{equation}
\begin{eqnarray}\label{eq:CS-E2}
p_1v_1\frac{\cos\alpha_+}{\sin\alpha_2} +  p_2v_2 =
(w_2\Gamma_2^2 - p_2)\left(\frac{v_1}{\sin\alpha_1} - v_2\right)\quad\quad\ \ 
\\ \nonumber
 - (w_1\Gamma_1^2-p_1)\frac{v_1}{\sin\alpha_2}\left(\frac{\sin\alpha_2}{\sin\alpha_1}+\cos\alpha_+\right)\,,
\end{eqnarray}
\begin{eqnarray}\nonumber
p_2-p_1 &=& 
\frac{\beta_1}{\cos\alpha_2}\left(\frac{\sin\alpha_2}{\sin\alpha_1}+\cos\alpha_+\right)w_1\Gamma_1u_1\cos\alpha_1
\\ \label{eq:CS-Px2}
&=& w_1u_1^2\cos^2\!\alpha_1\left(1+\frac{\tan\alpha_2}{\tan\alpha_1}\right)
\;,
\end{eqnarray}
\begin{eqnarray}\nonumber
(p_2-p_1)\tan\alpha_2 = w_2\Gamma_2u_2\left(\frac{\beta_1}{\sin\alpha_1}\!-\!\beta_2\right)\tan\alpha_2\hspace{0.1\textwidth}
\\ \label{eq:CS-Py2}
- w_1u_1^2\frac{\sin\alpha_1}{\cos\alpha_2}\left(\frac{\sin\alpha_2}{\sin\alpha_1}+\cos\alpha_+\right)\ ,\quad\quad\ 
\end{eqnarray}

The above set of 4 equations for the mass 
(Eq.~(\ref{eq:CS-M2})), energy  
(Eq.~(\ref{eq:CS-E2})), $x$-momentum  
(Eq.~(\ref{eq:CS-Px2})) 
and  $y$-momentum  
(Eq.~(\ref{eq:CS-Py2})),
together with the equation of state,
\begin{equation}
p_2 = (\hat{\gamma}_2-1)e_{\rm int,2} = (\hat{\gamma}_2-1)(e_2-\rho_2c^2)\ ,
\end{equation}
provide 5 equations for the 5 unknowns 
($e_2$, $p_2$, $\rho_2$, $v_2=\beta_2c$, $\alpha_2$), which can be solved in terms of the 5 knowns ($e_1$, $p_1$, $\rho_1$, $v_1=\beta_1c$, $\alpha_1$). 
In the simple case of a cold initial shell the latter reduce to $e_1=\rho_1c^2$, $p_1=0$.

\section{A 1D shock -- some relevant properties}
\label{sec:prev-res}

Let us start by considering the simpler one-dimensional shock problem, which in our context corresponds to the limit $\alpha_1\to0$ and therefore $\beta_p=\beta_1/\sin\alpha_1\to\infty$ for any fixed $u_1$.
Let us denote the upstream and downstream regions by 1 and 2, respectively. The initial free parameters are $\rho_1$, $p_1$, $e_1$, and $v_{12}$ (the relative velocity of regions 1 and 2). As long as $v_{12}>c_{s1}$ a shock is formed, and its velocity $v_{s1}$ relative to region 1 along with the conditions in region 2 ($\rho_2$, $p_2$, $e_2$) may be obtained by solving the 1D shock tube problem, i.e. the shock jump conditions along with the equation of state in region 2, where in region $i$ (where $i=1,\,2$),
\begin{equation}\label{eq:EoS-gamma}
p_i = (\hat{\gamma}_i-1)e_{\rm int,i} = (\hat{\gamma}_i-1)(e_i-\rho_ic^2)\ .
\end{equation}
Denoting $\Theta\equiv p/\rho c^2$, $\Theta_i=p_i/\rho_ic^2$ and $\hat{\gamma}_i=\hat{\gamma}(\Theta_i)$, we will use here the Taub-Matthews equation of state\citep{Ryu+06,Mignone+2007},
\begin{eqnarray}\nonumber
h(\Theta)&=&\frac {5}{2}\Theta+\sqrt{1+\frac{9}{4}\Theta^2}\ ,
\\ \label{eq:EoS-TM}
\hat{\gamma}(\Theta)&=&\frac{h(\Theta)-1}{h(\Theta)-1-\Theta}=\frac{1}{6}\left(8-3\Theta+\sqrt{4+9\Theta^2}\right)\ ,\quad
\\ \nonumber
\Theta &=& \frac{(5-3\hat{\gamma})(\hat{\gamma}-1)}{3\hat{\gamma}-4} = \frac{1}{8}\left(5h-\sqrt{16+9h^2}\right)\ ,
\end{eqnarray}
for which the dimensionless sound speed, $\beta_{c_s}=c_s/c$, is given by
\begin{equation}
\beta_{c_s}^2 = \frac{3\Theta^2+5\Theta\sqrt{\Theta^2+\frac{4}{9}}}{12\Theta^2+2+12\Theta\sqrt{\Theta^2+\frac{4}{9}}}\ .
\label{eq:Taub_soundspeed}
\end{equation}

The 1D shock jump conditions for mass and energy imply
\begin{equation}
\frac{e_2}{\rho_2} = \Gamma_{12}\frac{w_1}{\rho_1}-\frac{p_1}{\rho_2}\ \ \Longleftrightarrow\ \  
1+\frac{\Theta_2}{\hat{\gamma}_2-1}=\Gamma_{12}h_1-\Theta_1\frac{\rho_1}{\rho_2}\ ,
\end{equation}
which for a strong shock ($\Theta_1\ll\Theta_2\Leftrightarrow h_1-1\ll h_2-1$) reduces in the downstream region to $\Theta_2=(\hat{\gamma}_2-1)(\Gamma_{12}h_1-1)$. According to Eq.~(\ref{eq:EoS-TM}) this in turn implies
\begin{equation}\label{eq:gamma1a}
\hat{\gamma}_2 = \frac{4\Gamma_{12}h_1+1}{3\Gamma_{12}h_1}\;,
\end{equation}
such that the conditions in region 2 are given by \citep{blandford1976fluid}, which simplify further using Eq.~(\ref{eq:gamma1a}),
\begin{eqnarray}\nonumber
\frac{\rho_2}{\rho_1} &=& \frac{\Gamma_{12}[(4\Gamma_{12}+3)h_1+1]}{\Gamma_{12}h_1+1}\ ,
\\
\frac{p_2}{\rho_1c^2} &=& \frac{[(4\Gamma_{12}+3)h_1+1](h_1\Gamma_{12}-1)}{3h_1}\ ,\quad
\\ \nonumber
\frac{w_2}{\rho_1c^2} &=& \frac{[(4\Gamma_{12}+3)h_1+1](4\Gamma_{12}^2h_1^2-1)}{3h_1(\Gamma_{12}h_1+1)}\ ,
\\ \nonumber
u_{s1}^2 &=&
\frac{[(4\Gamma_{12}+3)h_1+1]^2\Gamma_{12}^2(\Gamma_{12}-1)}{1+\Gamma_{12}[2h_1(1+\Gamma_{12}[h_1(4\Gamma_{12}+5)-1])-1]}\ .
\end{eqnarray}

The self-consistency strong shock condition reads
\begin{equation}
1\ll\frac{h_2-1}{h_1-1}=\hat{\gamma}_2\frac{\Gamma_{12}h_1-1}{h_1-1}=\hat{\gamma}_2\left[1+\frac{h_1(\Gamma_{12}-1)}{h_1-1}\right]\ ,
\end{equation}
and it is always satisfied for a relativistic $\Gamma_{12}\gg1$, while for $\Gamma_{12}-1\lesssim1$ it corresponds to 
$\beta_{12}^2\gg p_1/\rho_1c^2$, which in all cases corresponds to a large Mach number
$\mathcal{M}=u_{s,1}/u_{c_s,1}\gg1$ where $u_{c_s,1}=(\beta_{c_s,1}^{-2}-1)^{-\frac{1}{2}}$
is the proper sound speed in region 1 and $\beta_{c_s,1}$ is the dimensionless sound speed.

Making a further assumption of an upstream that is not relativistically hot ($\Theta_1\ll1\Leftrightarrow h_1-1\ll1$), the expression for the adiabatic index reduces to
\begin{equation}\label{eq:gamma1}
\hat{\gamma}_2 = \frac{4\Gamma_{12}+1}{3\Gamma_{12}}\;.
\end{equation}

In order to later achieve a fully analytic solution for the 2D case, which is very useful, we will consider a cold region 1 where $p_1=0$, $e_1=w_1=\rho_1c^2$ and $h_1=1$, for which the shock is always strong in the sense that $\mathcal{M}\to\infty$ (not to be confused with the strong shock solution discussed below) and the conditions in region 2 are given by
\begin{eqnarray}\nonumber
\rho_2 &=& 4\Gamma_{12}\rho_1\;,\quad
p_2 = \frac{4}{3}u_{12}^2\rho_1c^2\;,\quad  \Theta_2 = \frac{\Gamma_{12}^2-1}{3\Gamma_{12}}\;, 
\\ \label{eq:1Dshock}
e_2&=&4\Gamma_{12}^2\rho_1c^2\,,\qquad
e_{\rm int,2} = \frac{4\Gamma_{12}u_{12}^2}{\Gamma_{12}+1}\rho_1c^2\ ,
\\ \nonumber
w_2 &=& 4\Gamma_{12}^2\left(1+\frac{\beta_{12}^2}{3}\right)\rho_1c^2\ ,
\\ \nonumber
\beta_{1,s}\!&=&\!\frac{4\Gamma_{12}u_{12}}{4\Gamma_{12}^2-1}\ ,\ \ 
u_{1,s} = \frac{4\Gamma_{12}u_{12}}{\sqrt{8\Gamma_{12}^2+1}}\ ,\ \ 
\Gamma_{s,1} = \frac{4\Gamma_{12}^2-1}{\sqrt{8\Gamma_{12}^2+1}}\ .
\end{eqnarray}
Similarly
\begin{eqnarray}
u_{12} &=& \frac{1}{2}\sqrt{u_{s,1}^2-2+\sqrt{4+5u_{s,1}^2+u_{s,1}^4}}\ ,
\\ \label{eq:1Dbeta_sh}
\beta_{2,s} &=& -\beta_{s,2} =-\frac{\beta_{s,1}-\beta_{21}}{1-\beta_{21}\beta_{s,1}} = \frac{\beta_{1,s}-\beta_{12}}{1-\beta_{12}\beta_{1,s}}
= \frac{\beta_{12}}{3}\ ,
\end{eqnarray}
where Eq.~(\ref{eq:1Dbeta_sh}) means that for our equation of state, in the rest frame of the downstream fluid (region 2), the speed at which the shock is receding is a third of the incoming upstream speed.

\section{Solving the 2D Cold Shell Equations}
\label{sec:solutions}

Note that  
\begin{equation}\label{eq:Gamma12}
\Gamma_{12}=\Gamma_1\Gamma_2(1-\vec{\beta}_1\cdot\vec{\beta}_2)=\Gamma_1\Gamma_2(1-\beta_1\beta_2\sin\alpha_1)\;,   
\end{equation}
which can be used together\footnote{Note that the quantities in equations (\ref{eq:gamma1}) and (\ref{eq:1Dshock}) are expressed in terms of the relative velocity or Lorentz factors of regions 1 and 2, which may in turn be evaluated in any frame.} with Eq.~(\ref{eq:gamma1}) to express the adiabatic index in region 2 as
\begin{equation}
\hat{\gamma}_2 = \frac{4\Gamma_{12}+1}{3\Gamma_{12}} =
\frac{4}{3}+[3\Gamma_1\Gamma_2(1-\beta_1\beta_2\sin\alpha_1)]^{-1}\;.
\end{equation}

For a cold shell Eqs.~(\ref{eq:CS-M2})-(\ref{eq:CS-Py2}) reduce to 
\begin{eqnarray}\label{eq:CS-M2c}
\frac{\Gamma_2\rho_2}{\Gamma_1\rho_1} &=& \frac{-\dot{\mathbb{V}}_1}{\dot{\mathbb{V}}_2}=\frac{1+\frac{\sin\alpha_1}{\sin\alpha_2}\cos\alpha_+}{1-\frac{\beta_2}{\beta_1}\sin\alpha_1}\ ,
\\ \label{eq:CS-E2c}
p_2\beta_2 &=&
(w_2\Gamma_2^2 - p_2)\left(\frac{\beta_1}{\sin\alpha_1} - \beta_2\right)\quad\quad\ \ 
\\ \nonumber
& & -\frac{\rho_1c^2\Gamma_1u_1}{\sin\alpha_2}\left(\frac{\sin\alpha_2}{\sin\alpha_1}+\cos\alpha_+\right)\,,
\\ \label{eq:CS-Px2c}
p_2 &=&\rho_1c^2u_1^2\cos^2\!\alpha_1\left(1+\frac{\tan\alpha_2}{\tan\alpha_1}\right)
\;,
\\ \nonumber
p_2 &=& w_2\Gamma_2u_2\left(\frac{\beta_1}{\sin\alpha_1}\!-\!\beta_2\right)
\\ \label{eq:CS-Py2c}
& & - \rho_1c^2u_1^2\left(1+\frac{\sin\alpha_1}{\sin\alpha_2}\cos\alpha_+\right)\ ,\quad\quad\ 
\end{eqnarray}

\begin{figure}
\centering
\includegraphics[trim={0cm 0cm 0cm 0cm},clip,scale=0.6,width=0.48\textwidth]{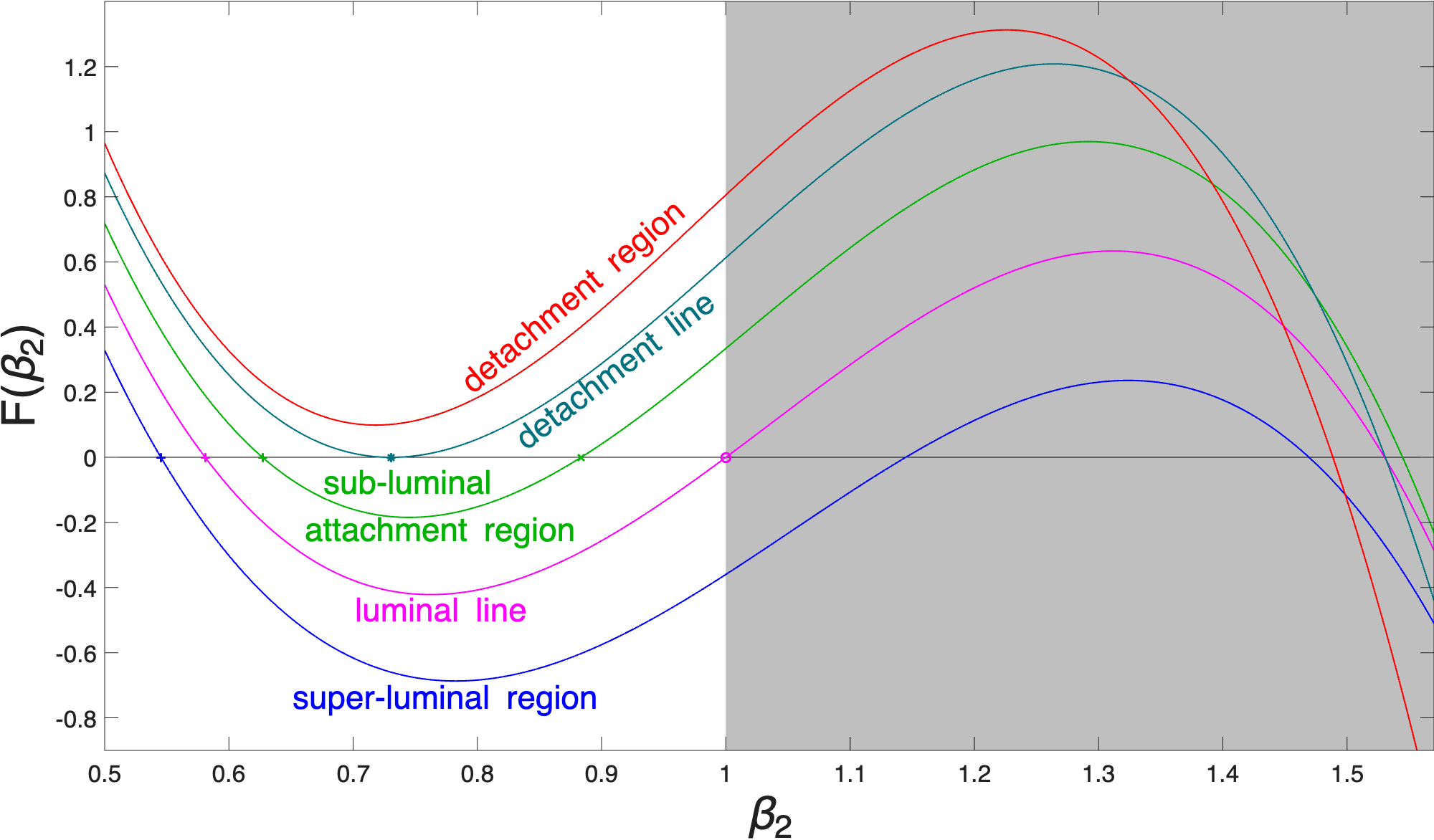}
\vspace{-0.5cm}
\caption{The roots of Eq.~(\ref{eq:roots}) for $u_1=1$ and different values of $\alpha_1$ corresponding to the different regions and critical lines in the $u_1$-$\alpha_1$ parameter space: 1. The super-luminal region where $0\leq\alpha_1<\alpha_{\rm lum}$ (\blue{blue line}; $\alpha_1=0.73$), 2. The luminal line ($\sin\alpha_{\rm lum}=\beta_1=\frac{1}{\sqrt{2}}$; $\alpha_1=\alpha_{\rm lum}=\frac{\pi}{4}$; \magenta{magenta line}), 3. The sub-luminal attachment region where  $\alpha_{\rm lum}<\alpha_1\leq\alpha_{\rm det}$ ($\alpha_1=0.856$; \textcolor[rgb]{0,0.7,0}{green line}), 4. The detachment (or sonic) line ($\alpha_1=\alpha_{\rm det}=0.94953116034$; \textcolor[rgb]{0,0.4431,0.4980}{turquoise line}), 5. The detached region where $\alpha_1>\alpha_{\rm det}$ ($\alpha_1=1.1$; \red{red line}). The gray shaded area where $\beta_2\geq1$ is unphysical.}
\label{fig:roots}
\end{figure}

Combining Eq.~(\ref{eq:1Dshock}) with Eqs.~(\ref{eq:CS-M2c}) and (\ref{eq:CS-Px2c}), respectively, gives
\begin{eqnarray}
\frac{\Gamma_2}{\Gamma_1}4\Gamma_{12} &=& 
\frac{1+\frac{\sin\alpha_1}{\sin\alpha_2}\cos\alpha_+}{1-\frac{\beta_2}{\beta_1}\sin\alpha_1}\ ,
\\ \label{eq:u12}
\frac{4}{3}u_{12}^2 &=& u_1^2\cos^2\!\alpha_1\left(1+\frac{\tan\alpha_2}{\tan\alpha_1}\right)\ ,
\end{eqnarray}
which together with Eq.~(\ref{eq:Gamma12})
leads to a singe equation for $\beta_2$.
This is a fourth order polynomial that has a simple real root, $\beta_2=\beta_1\sin\alpha_1$, which is nonetheless not physical (as it doesn't satisfy, e.g. the y-momentum equation). Using this fact leaves us with the third order polynomial,
\begin{equation}\label{eq:roots}
F(\beta_2) = A\beta_2^3+B\beta_2^2+C\beta_2+D=0\ ,
\end{equation}
with the following coefficients,
\begin{eqnarray}\nonumber
A &=& -4\beta_1\left[5-\beta_1^2-(3+\beta_1^2)\cos(2\alpha_1)\right]\ ,
\\ \nonumber
B &=& 16\left[2+3\beta_1^2-\beta_1^2\cos(2\alpha_1)\right]\sin\alpha_1\ ,
\\ \label{eq:coef}
C &=& -4\beta_1\left[11+5\beta_1^2-(5+3\beta_1^2)\cos(2\alpha_1)\right]\ ,
\\ \nonumber
D &=& 32\beta_1^2\sin\alpha_1\ .
\end{eqnarray}
The three roots of this equation, corresponding to $k=1,2,3$, are given by
\begin{eqnarray}\nonumber
\beta_{2,k} &=& \frac{1}{3A}\left(F_k+\frac{B^2-3AC}{F_k}-B\right)\ ,
\\
F_k &=& z_k\left(\frac{G+\sqrt{4(3AC-B^2)^3+G^2}}{2}\right)^{1/3}\ ,
\\ \nonumber
G &=& 9ABC-2B^3-27A^2D\ ,
\\ \nonumber
z_k &=& \exp\left[i\,2\pi\frac{1-k}{3}\right] \Longleftrightarrow\ z_1 = 1\,,\ \  z_2=\bar{z}_3=-\frac{1+\sqrt{3}\,i}{2}\ ,\ \ 
\end{eqnarray}
where $z_k$ are the three roots of the complex equation $z^3=1$. The first root, $k=1$ (denoted by a `+' symbol in Fig.~\ref{fig:roots}), corresponds to the weak shock solution\footnote{In both solutions the shock is technically strong in the sense discussed in \S\,\ref{sec:prev-res}, and their label as weak or strong refers to their strength relative to each other, in analogy to the two analogous solutions in regular shock reflection.} (denoted with a subscript `w', i.e. $\beta_{2,1}\to\beta_{2,w}$ etc.) and exists for $0\leq\alpha_1\leq\alpha_{\rm det}(u_1)$ where $\alpha_{\rm det}(u_1)$ is the detachment angle for a given value of $u_1$,
i.e. in the attachment regions. The second root, $k=2$ (denoted by an `x' symbol in Fig.~\ref{fig:roots}), corresponds to the strong shock solution (denoted with a subscript `s', i.e. $\beta_{2,2}\to\beta_{2,s}$ etc.) and exists for $\beta_1=\sin\alpha_{\rm lum}(u_1)<\sin\alpha_1\leq\sin\alpha_{\rm det}(u_1)$, i.e. in the sub-luminal attachment region between the luminal and detachment lines. The third root, $k=3$, is unphysical (always being either complex or larger than 1 when real). Along the luminal line $1=\beta_p=\beta_1/\sin\alpha_1$ and substituting $\sin\alpha_1\to\beta_1$ in the expressions for the coefficients A-D (Eq.~(\ref{eq:coef})) gives
\begin{eqnarray}\nonumber
A&\to&-8\beta_1(1+\beta_1^2)^2\ ,
\\ \label{eq:luminal-line-strong}
B&\to&32\beta_1(1+\beta_1^2+\beta_1^4)\ ,
\\ \nonumber
C&\to&-24\beta_1(1+\beta_1^2)^2\ ,
\\  \nonumber
D&\to&32\beta_1^3\ ,  
\end{eqnarray}
and their sum vanishes ($A+B+C+D=0$) such that the $k=2$ root is always 1 along the luminal line (see the magenta circle in Fig~\ref{fig:roots}).
This signifies the divergence of $\Gamma_{2,s}$ for the strong shock solution as it approaches the luminal line from the sub-luminal side. For this reason it cannot cross the luminal line and is confined to the sub-luminal region. The strong and weak shock solutions coincide at the detachment line, which bounds both of them.

Substituting the above two analytic solutions for $\beta_2$ into Eq.~(\ref{eq:Gamma12}) now provides the corresponding values of $\Gamma_{12}$, which can in turn be substituted into Eq.~(\ref{eq:1Dshock}) to analytically find 
the corresponding values of the remaining hydrodynamic variables in region 2. Finally, the corresponding angle $\alpha_2$ may be obtained through Eq.~(\ref{eq:CS-Px2c}),
\begin{equation}
\tan\alpha_2 = \left[\frac{4(\Gamma_{12}^2-1)}{3u_1^2\cos^2\!\alpha_1}-1\right]\tan\alpha_1\ .
\end{equation}

\section{Results}
\label{sec:results}

\subsection{The Weak Shock Solution}

\begin{figure}
\centering
\includegraphics[trim={0cm 0cm 0cm 0cm},clip,scale=0.6,width=0.48\textwidth]{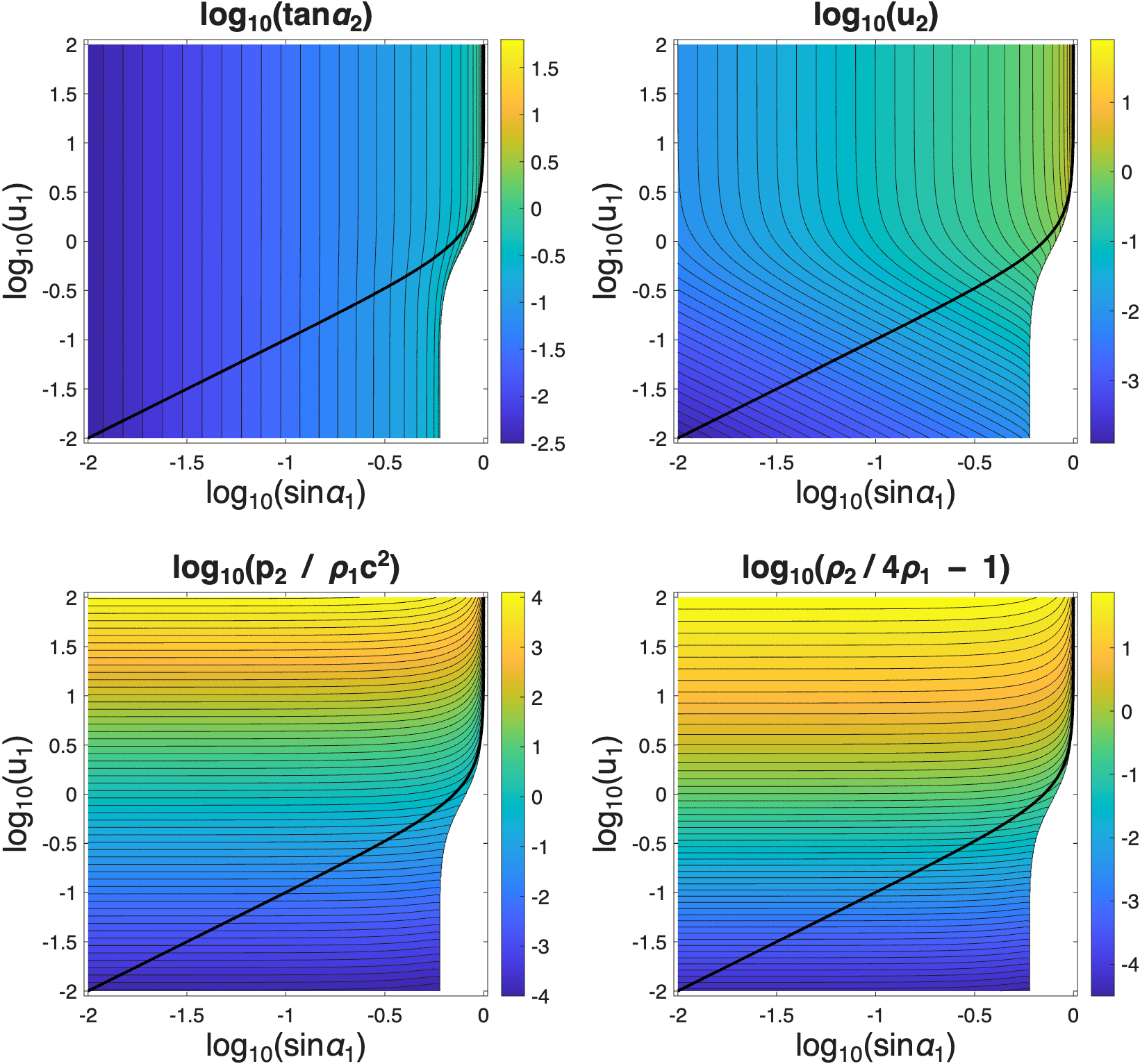}
\vspace{-0.5cm}
\caption{Contour plots of the hydrodynamic variables in the shocked portion (region 2) of a cold shell (region 1) colliding with a wall at an incidence angle $\alpha_1$ and proper velocity $u_1$ normal to the shell's front edge, shown in the $\log_{10}(\sin\alpha_1)$\,--\,$\log_{10}(u_1)$ plane. Shown are the tangent of the angle $\alpha_2$ of the shock relative to the wall (\textit{top left panel}), the proper velocity (\textit{top right panel}), the normalized pressure (\textit{bottom left panel}) and the normalized proper density (\textit{bottom right panel}). The thick solid black line is the luminal line (where $v_p = v_1/\sin\alpha_1 = 1\Leftrightarrow u_1=\tan\alpha_1$). The white region in the bottom right corner is the detachment region where there is no regular solution (see Fig.~\ref{fig:critical_lines}).}
\label{fig:results_sin}
\end{figure}

\begin{figure}
\centering
\includegraphics[trim={0cm 0cm 0cm 0cm},clip,scale=0.6,width=0.48\textwidth]{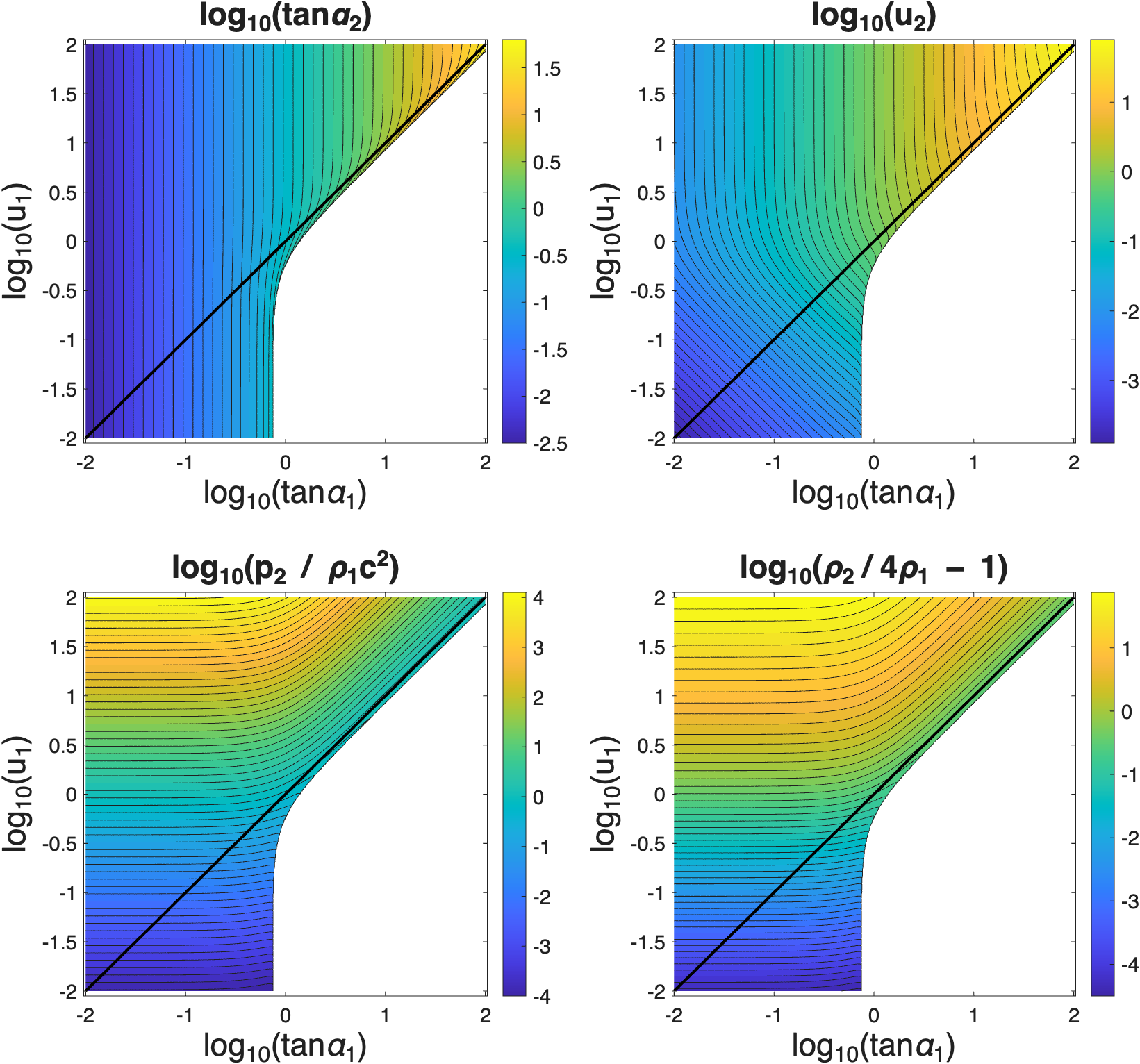}
\vspace{-0.5cm}
\caption{Same as Fig.~\ref{fig:results_sin} but shown in the $\log_{10}(\tan\alpha_1)$\,--\,$\log_{10}(u_1)$ plane.}
\label{fig:results_tan}
\end{figure}

\begin{figure}
\centering
\includegraphics[trim={0cm 0cm 0cm 0cm},clip,scale=0.6,width=0.48\textwidth]{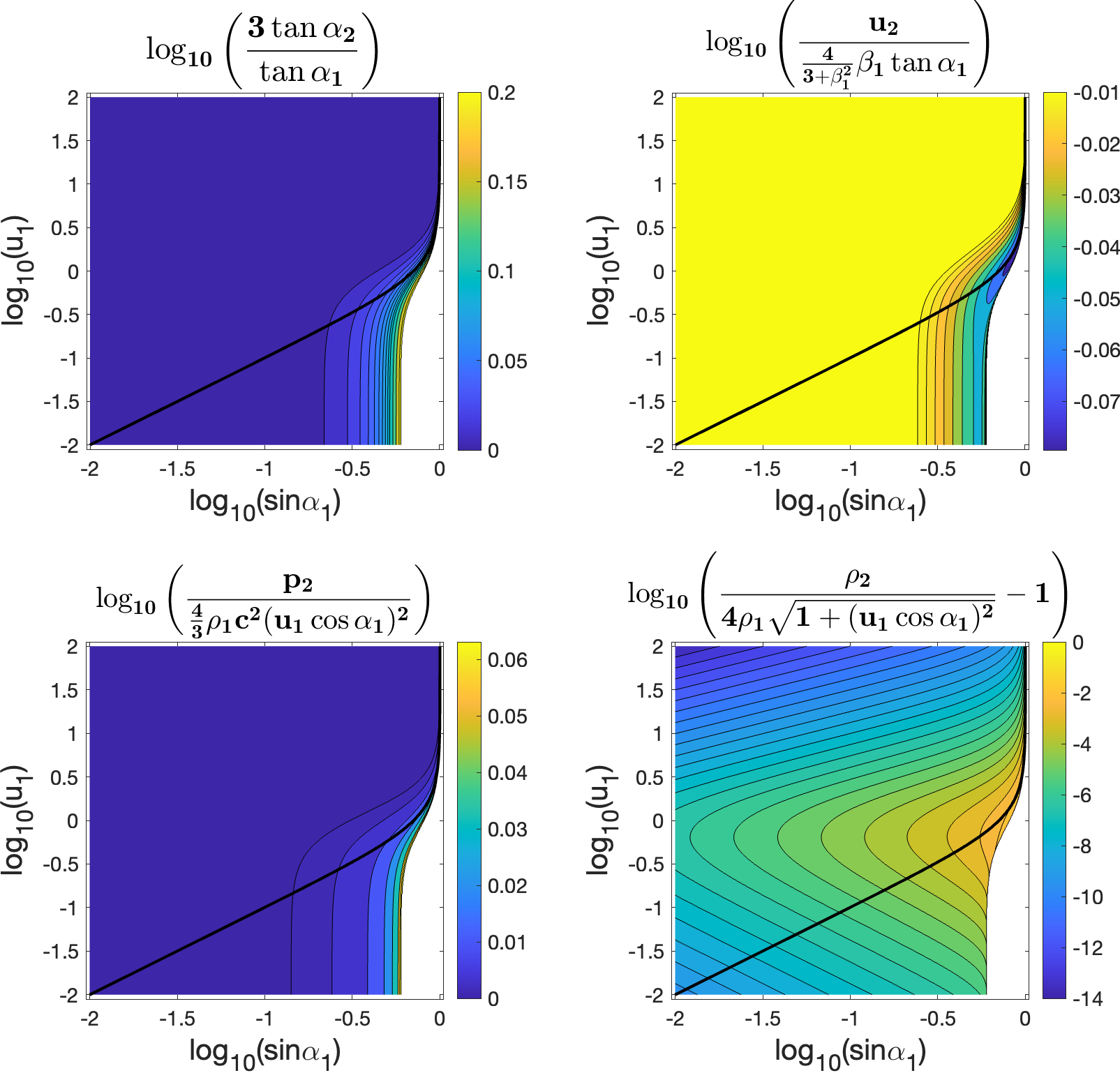}
\vspace{-0.5cm}
\caption{Similar to Fig.~\ref{fig:results_sin} but with each of the quantities normalized to a simple analytic function that captures most of its variation within this parameter space. For the density (\textit{bottom right panel}) we show the log of the the difference of this ration from unity, as it is a particularly good approximation everywhere.}
\label{fig:results_norm}
\end{figure}

The results for the weak shock solution for $\alpha_2$, $u_2$, $p_2$ and $\rho_2$ are shown in Figures~\ref{fig:results_sin} and \ref{fig:results_tan}, 
in the 
$\log_{10}(\sin\alpha_1)$\,--\,$\log_{10}(u_1)$ plane and in the $\log_{10}(\tan\alpha_1)$\,--\,$\log_{10}(u_1)$ plane, 
respectively. Fig.~\ref{fig:results_norm} 
shows in the 
$\log_{10}(\sin\alpha_1)$\,--\,$\log_{10}(u_1)$ plane each of the quantities $\alpha_2$, $u_2$, $p_2$ and $\rho_2$ normalized to a simple analytic function that captures most of its variation within this parameter space.

For $\alpha_1\to0$ the 2D case approaches the 1D case, and
from comparing the relevant limit of the particle number equation (\ref{eq:CS-M2c}) (e.g. $\beta_2=\mathcal{O}(\alpha_1)$, $\Gamma_2=1+\mathcal{O}(\alpha_1^2)$, $\Gamma_{12}=\Gamma_1[1+\mathcal{O}(\alpha_1^2)]$, $\cos\alpha_+=1-\mathcal{O}(\alpha_1^2)$ where in 1D $\frac{\rho_2}{\rho_1}=\Gamma_{12}(1+\frac{\beta_{12}}{\beta_{s}})$), one obtains that
\begin{equation}\label{eq:alpha12_ratio}
\frac{\sin\alpha_1}{\sin\alpha_2}
\xrightarrow[\alpha_1\to0]{}
\frac{\beta_{12}}{\beta_{s}}\xrightarrow[\text{strong\;shock}]{}3\ ,
\end{equation}
which has the limiting value of 3 for a strong shock, both in the Newtonian and in the ultra-relativistic regimes. From comparing the $x$-momentum equation a similar result is obtained for $\tan\alpha_1/\tan\alpha_2$, which is to be expected as in this limit $\tan\alpha_1/\tan\alpha_2\approx \sin\alpha_1/\sin\alpha_2\approx\alpha_1/\alpha_2$ since $\alpha_2<\alpha_1\ll1$. This result can nicely be seen from the upper-left panel of Fig.~\ref{fig:results_norm}.
The approximate expressions for the pressure and density in region 2 (bottom left and right panels of Fig.~\ref{fig:results_norm}, respectively) simply use the component of the pre-collision proper speed normal to the wall ($u_1\cos\alpha_1$) and the corresponding Lorentz factor, respectively, and naturally work better for smaller incidence angles $\alpha_1$ (since Eq.~(\ref{eq:u12}) together with Eq.~(\ref{eq:alpha12_ratio}) imply $u_{12}\to u_1\cos\alpha_1$ for $\alpha_1\to0$).

\subsection{The Parameter Space}

\begin{figure}
\centering
\includegraphics[trim={0cm 0cm 0cm 0cm},clip,scale=1,width=0.475\textwidth]{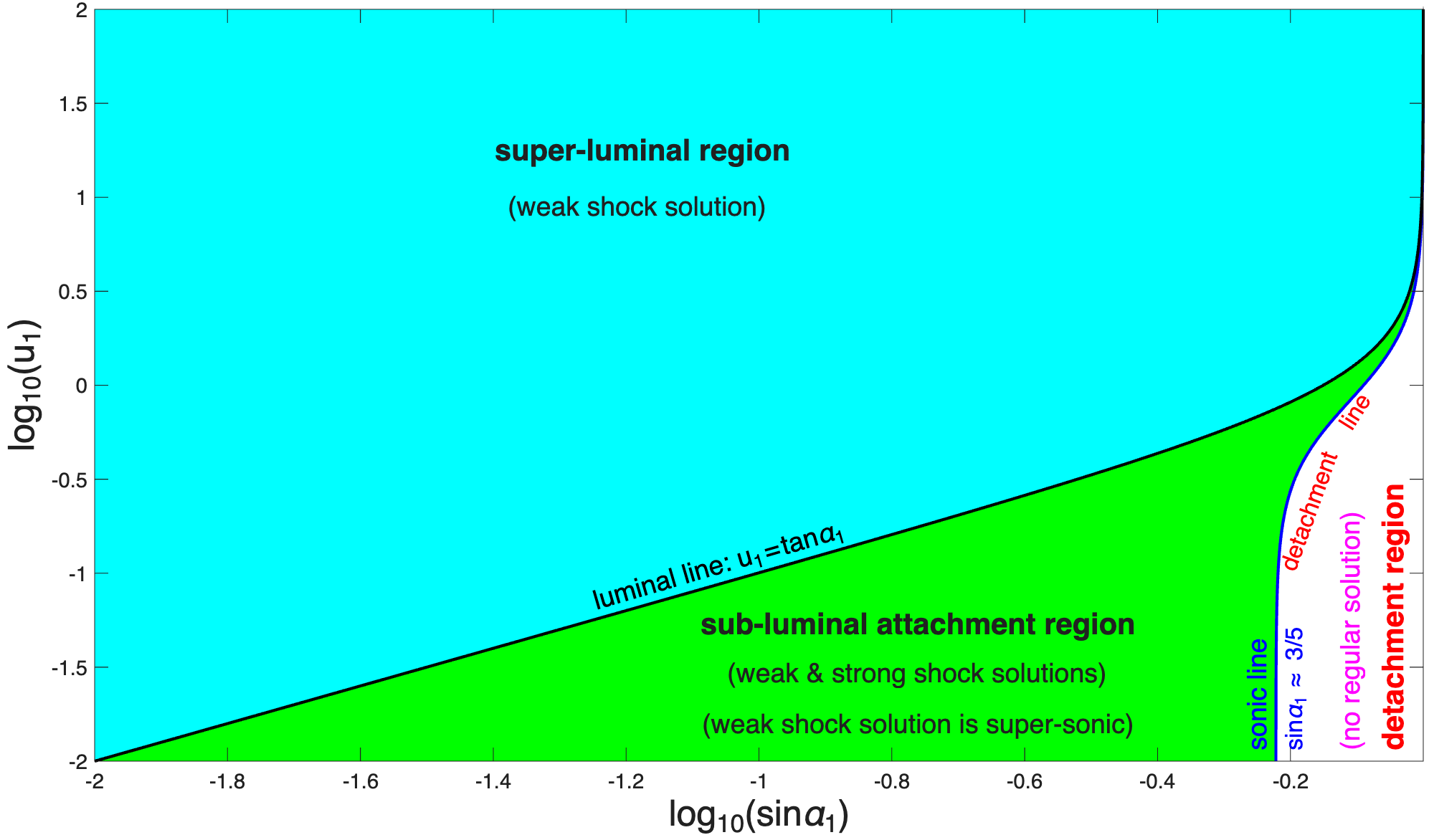}
\vspace{0.1cm}\\
\includegraphics[trim={0cm 0cm 0cm 0cm},clip,scale=1,width=0.475\textwidth]{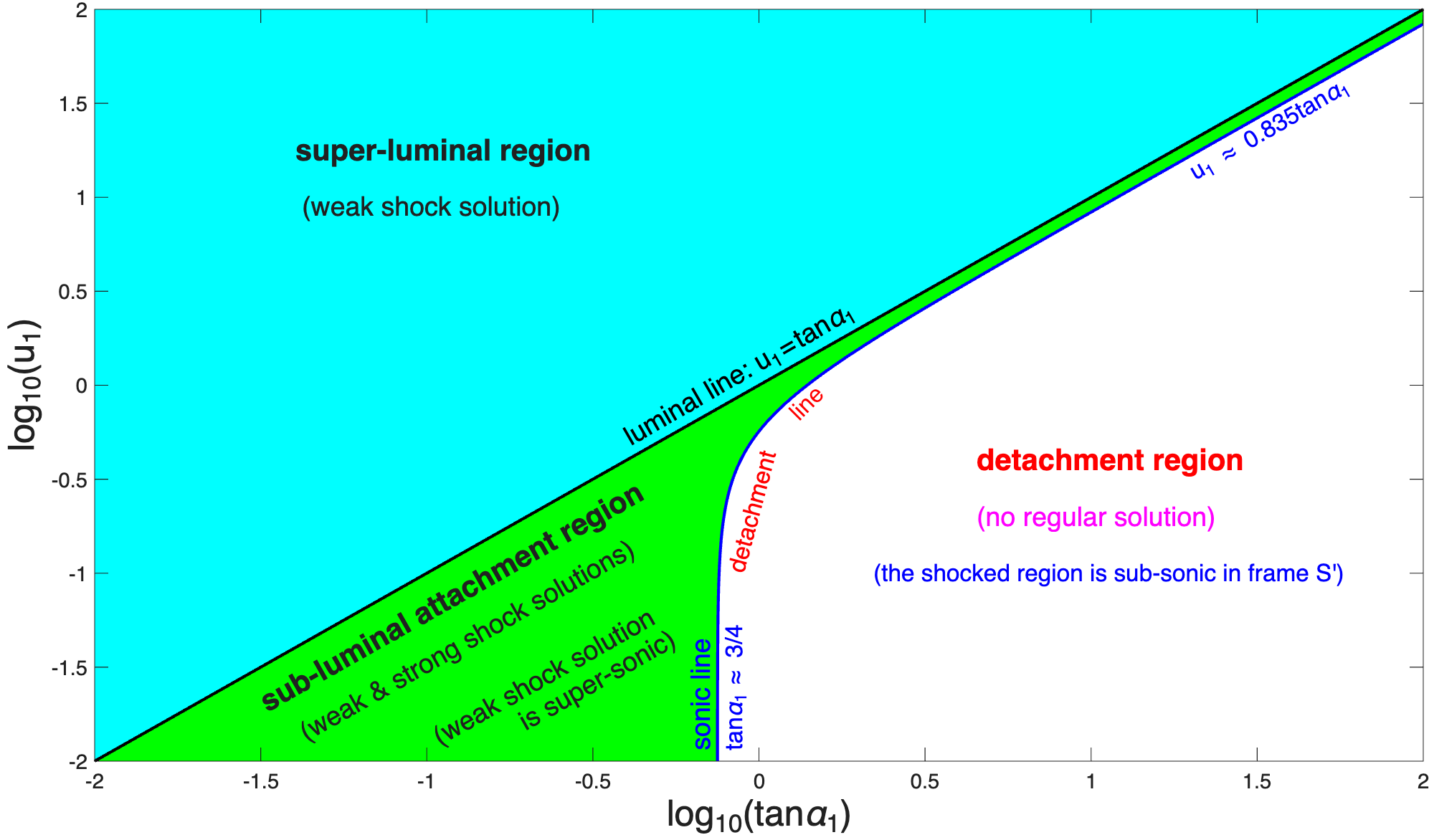}
\vspace{-0.1cm}
\caption{Maps of the $u_1$-$\alpha_1$ parameter space displaying the three different regions separated by two critical lines, shown in the $\log_{10}(\sin\alpha_1)$\,--\,$\log_{10}(u_1)$ plane (\textit{top panel}) and the $\log_{10}(\tan\alpha_1)$\,--\,$\log_{10}(u_1)$ plane (\textit{bottom panel}).
Above the luminal line (solid black line, where $v_p = v_1/\sin\alpha_1 = c\Leftrightarrow u_1=\tan\alpha_1$) is the super-luminal region (shaded in \cyan{cyan}) where only the weak shock solution exists. Between the luminal line and the sonic/detachment line (solid \blue{blue} line where $\beta'_2=(\beta_p-\beta_2)/(1-\beta_p\beta_2)=c_{s,2}\Leftrightarrow\beta_p=(\beta_2+c_{s,2})/(1+\beta_2c_{s,2})$) is the sub-luminal attachment region (shaded in \dgreen{green}) where both the weak and strong shock solutions exist, and their region 2 is  super-sonic and sub-sonic, respectively, in frame $S'$. At the sonic/detachment line the two solutions coincide, and below it is the detachment region (in white) where the shocked region becomes sub-sonic in frame $S'$,
point $P$ detaches from the wall and shocked fluid spills into the vacuum region.}
\label{fig:critical_lines}
\end{figure}

Figure~\ref{fig:critical_lines} shows maps of the relevant $u_1$-$\alpha_1$ parameter space that show how it divides into three distinct regions by two critical lines. The top and bottom panels show the $\log_{10}(\sin\alpha_1)$\,--\,$\log_{10}(u_1)$ plane and the $\log_{10}(\tan\alpha_1)$\,--\,$\log_{10}(u_1)$ plane, respectively.
The \textbf{\textit{luminal line}} (solid black line) corresponding to $v_p = v_1/\sin\alpha_1 = c\Leftrightarrow u_1=\tan\alpha_1$ divides between the super-luminal region (shaded in \cyan{cyan}) above it, and the sub-luminal attachment region below it (shaded in \dgreen{green}). The sonic line (solid \blue{blue} line) corresponding to $\beta'_2=(\beta_p-\beta_2)/(1-\beta_p\beta_2)=c_{s,2}\Leftrightarrow\beta_p=(\beta_2+c_{s,2})/(1+\beta_2c_{s,2})$), which coincides with the detachment line for a cold region~1, divides between the sub-luminal attachment region above it (where region 2 of the weak shock solution is super-sonic in frame $S'$) and the
detachment region region below it (in white) where the shocked region becomes sub-sonic in frame $S'$. In the detachment region there is no regular solution (i.e. a solution where the interaction point $P$ is attached to the wall) and instead point $P$ detaches from the wall and high-pressure shocked material (from region 2) spills into the vacuum (region 0). This can occur, e.g., when a large high-velocity meteorite impacts a solid planet, melting the rock at the impact point and melted rock fragments are ejected from the impact location to the surrounding region (where small enough fragments may re-solidify in flight). Note that a 1D planar collision corresponds to the limit $\alpha_1\to0$ and therefore $\beta_p=\beta_1/\sin\alpha_1\to\infty$ for any fixed $u_1$, i.e. it is always in the super-luminal regime where a singe solution exists. While this single solution is called a weak shock solution, the shock itself can still be very strong in terms of its Mach number ($\mathcal{M}\gg1$ and in our case $\mathcal{M}\to\infty$).

\begin{figure}
\centering
\includegraphics[trim={0cm 0cm 0cm 0cm},clip,scale=1,width=0.482\textwidth]{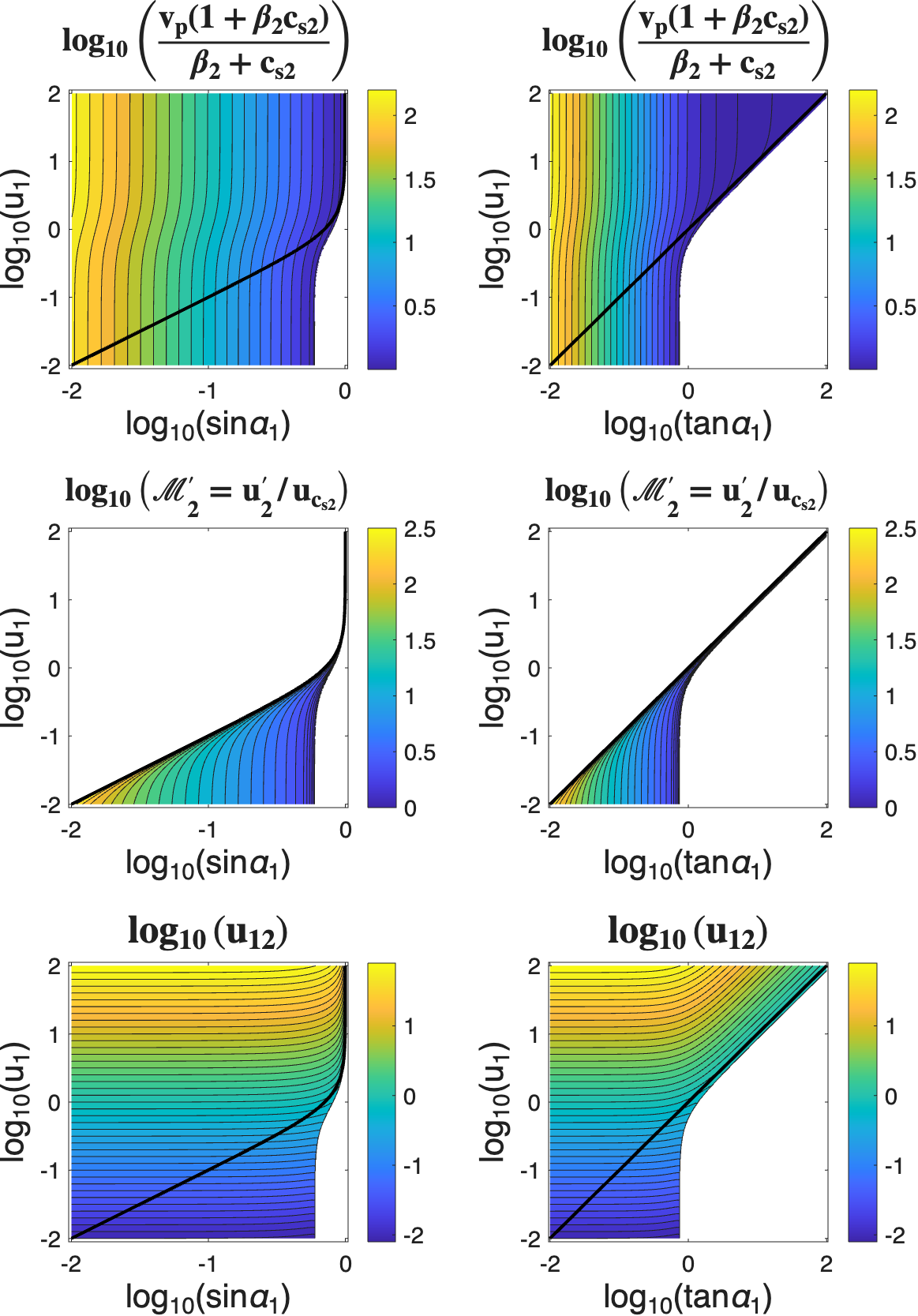}
\vspace{-0.5cm}
\caption{The \textit{top panels} show the ratio of the velocity of point $P$ ($v_p$) and that of a sound wave in region 2 in the lab frame. The thick solid black line is the luminal line (where $v_p = v_1/\sin\alpha_1 = 1\Leftrightarrow u_1=\tan\alpha_1$).
The \textit{middle panels} show 
the Mach number of region 2 for the weak shock solution in the steady-state frame $S'$, i.e. $\mathcal{M}'_2=u'_2/u_{c_{s2}}$, which is defined only in the sub-luminal attachment region.
The \textit{left panels} show the $\log_{10}(\sin\alpha_1)$\,--\,$\log_{10}(u_1)$ plane, while the \textit{right panels} show the $\log_{10}(\tan\alpha_1)$\,--\,$\log_{10}(u_1)$ plane. All panels are for the weak shock solution.}
\label{fig:results_6panel}
\end{figure}

In the attachment regions there is always a regular solution for which we show our analytic results. In the sub-luminal attachment region (shaded in \dgreen{green}) there are two possible solutions -- the strong shock solution and weak shock solution.
In the super-luminal region (shaded in \cyan{cyan}) only the weak shock solution exists, as we explain below.

The \textit{top panels} of Fig.~\ref{fig:results_6panel} show  the ratio of $v_p$ and and the velocity of a sound wave in region 2 propagating parallel to the wall toward point $P$ for the weak shock solution, which clearly exceeds unity in the super-sonic region and equals unity along the sonic line.
The same can also be seen from the \textit{middle panels} of 
Fig.~\ref{fig:results_6panel} that show the Mach number of region 2 for the weak shock solution in the rest frame of point $P$ (frame $S'$), $\mathcal{M}'_2=u'_2/u_{c_{s2}}$, which is defined only in the sub-luminal attachment region. The \textit{bottom panels} of Fig.~\ref{fig:results_6panel} 
show the proper relative velocity of regions 1 and 2 which serves as a useful measure of the shock strength (see, e.g., Eq.~(\ref{eq:1Dshock})). The shock strength along the sonic line approaches a constant value in the relativistic limit ($u_1\gg1$).

\subsection{The Strong Shock Solution}

\begin{figure}
\centering
\includegraphics[trim={0cm 0cm 0cm 0cm},clip,scale=0.6,width=0.48\textwidth]{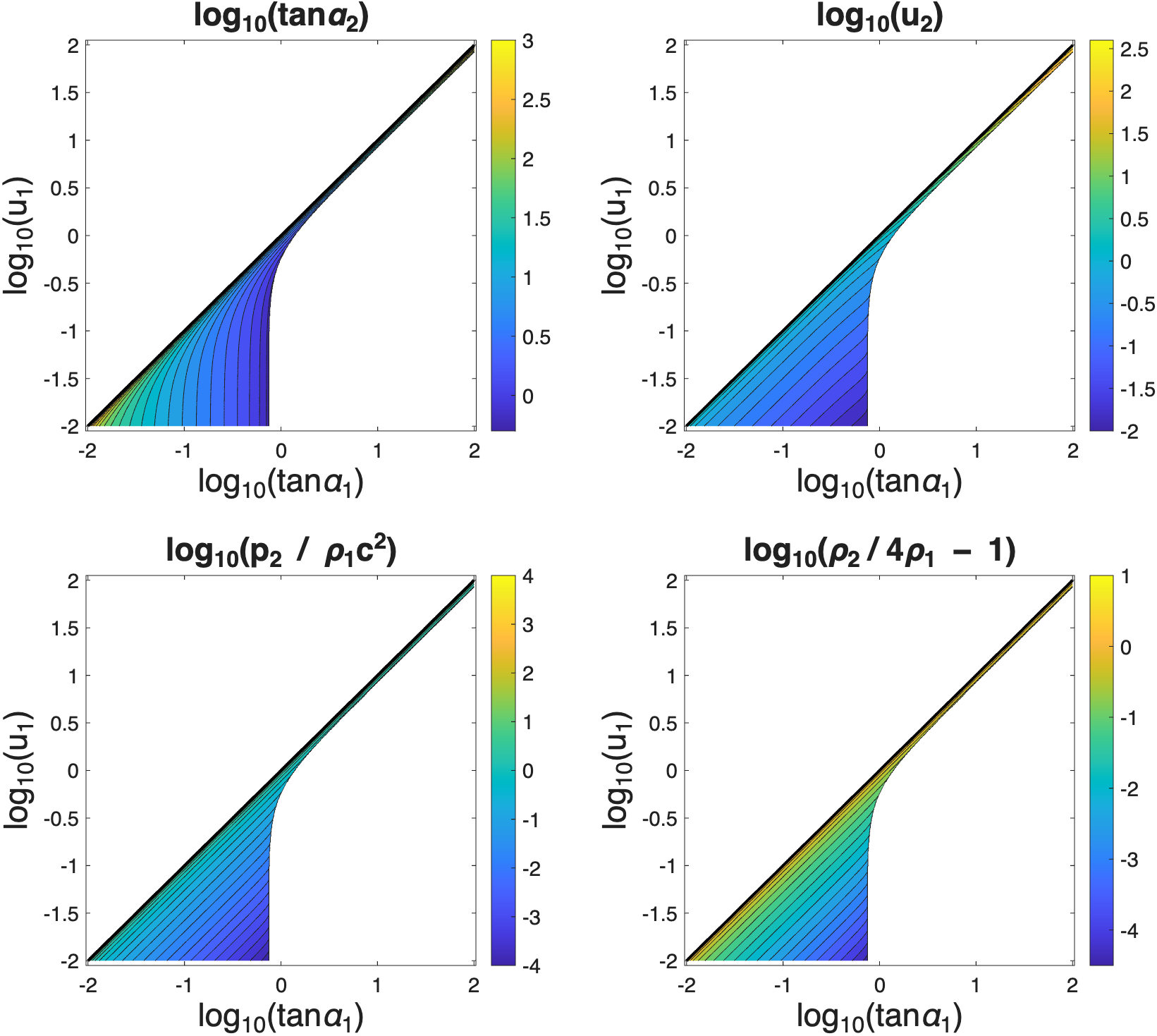}
\vspace{-0.5cm}
\caption{Same as Fig.~\ref{fig:results_tan} but for the strong shock solution. This solution exists only between the luminal line and the sonic/detachment line, i.e. the sub-luminal attachment region.}
\label{fig:results_tan_strong}
\end{figure}

Figure~\ref{fig:results_tan_strong} shows results for the strong shock solution, which exists only between the luminal line and the sonic/detachment line, i.e. the sub-luminal attachment region (which is shaded in \dgreen{green} in Fig.~\ref{fig:critical_lines}). In the strong shock solution region 2 is always sub-sonic in frame $S'$, i.e. in causal contact through sound waves with the point where it crossed the shock front,
in contrast to the weak shock solution.\footnote{A shocked fluid element in the uniform region 2 will always be in causal contact with part of the shock surface, since the shock downstream is always sub-sonic in the frame where the shock is stationary and normal.}

\begin{figure}
\centering
\includegraphics[trim={0cm 0cm 0cm 0cm},clip,scale=1,width=0.48\textwidth]{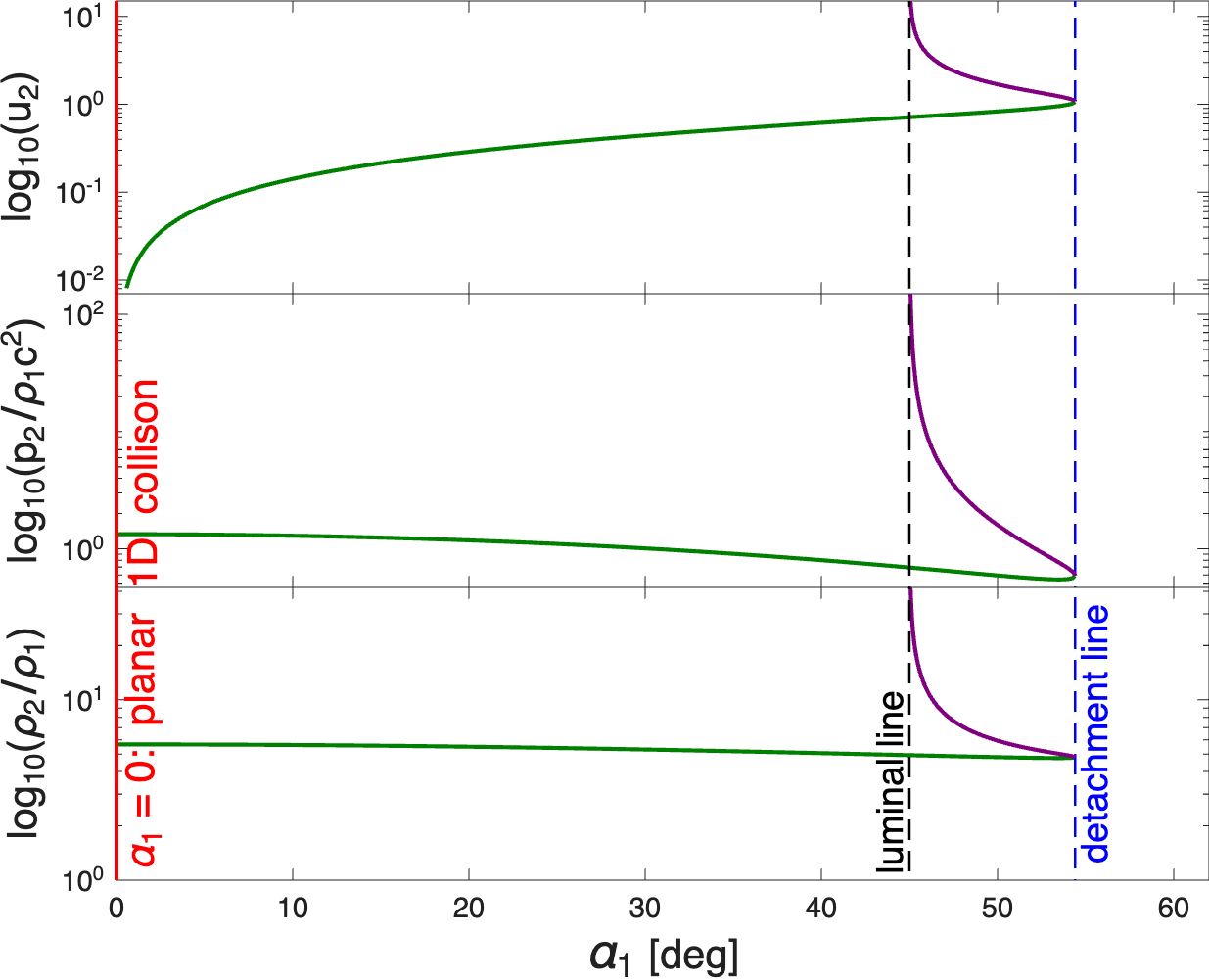}
\vspace{-0.45cm}
\caption{The proper speed, $u_2$, normalized pressure, $p_2/\rho_1c^2$, and normalized proper rest-mass density, $\rho_2/\rho_1$ (i.e. shock compression ration), in the shocked region 2, for $u_1=1$ as a function of the cold shell's shock incidence angle $\alpha_1$, for the weak (“W” in \textcolor[rgb]{0,0.5,0}{green}) and strong
(“S” in \textcolor[rgb]{0.5,0,0.5}{purple}) shock regular solutions. The strong shock solution exists only between the luminal and detachment lines, while the weak shock solution smoothly transitions across the luminal line, and is the only solution in the super-luminal regime, which always includes the 1D planar collision case corresponding to $\alpha_1=0$ (in \red{red}).}
\label{fig:SW_solutions_a1}
\end{figure}

Figure~\ref{fig:SW_solutions_a1} shows the proper speed, $u_2$, normalized pressure, $p_2/\rho_1c^2$, and normalized proper rest-mass density, $\rho_2/\rho_1$ (i.e. shock compression ration), in the shocked region 2, for $u_1=1$ as a function of the cold shell's shock incidence angle $\alpha_1$, for the weak (“W” in \textcolor[rgb]{0,0.5,0}{green}) and strong
(“S” in \textcolor[rgb]{0.5,0,0.5}{purple}) shock regular solutions. It can be seen that for the strong shock solution all of these hydrodynamic variables diverge towards the luminal line. This implies that the strong shock solution cannot cross the luminal line and does not exist in the super-luminal region. This was also explicitly shown in \S\,\ref{sec:solutions} for our analytic solution. This behavior is similar to that which we have found for relativistic shock reflection \citep{Granot-Rabinovich-24,Bera+24}. 

\begin{figure}
\centering
\includegraphics[trim={0cm 0cm 0cm 0cm},clip,scale=0.6,width=0.48\textwidth]{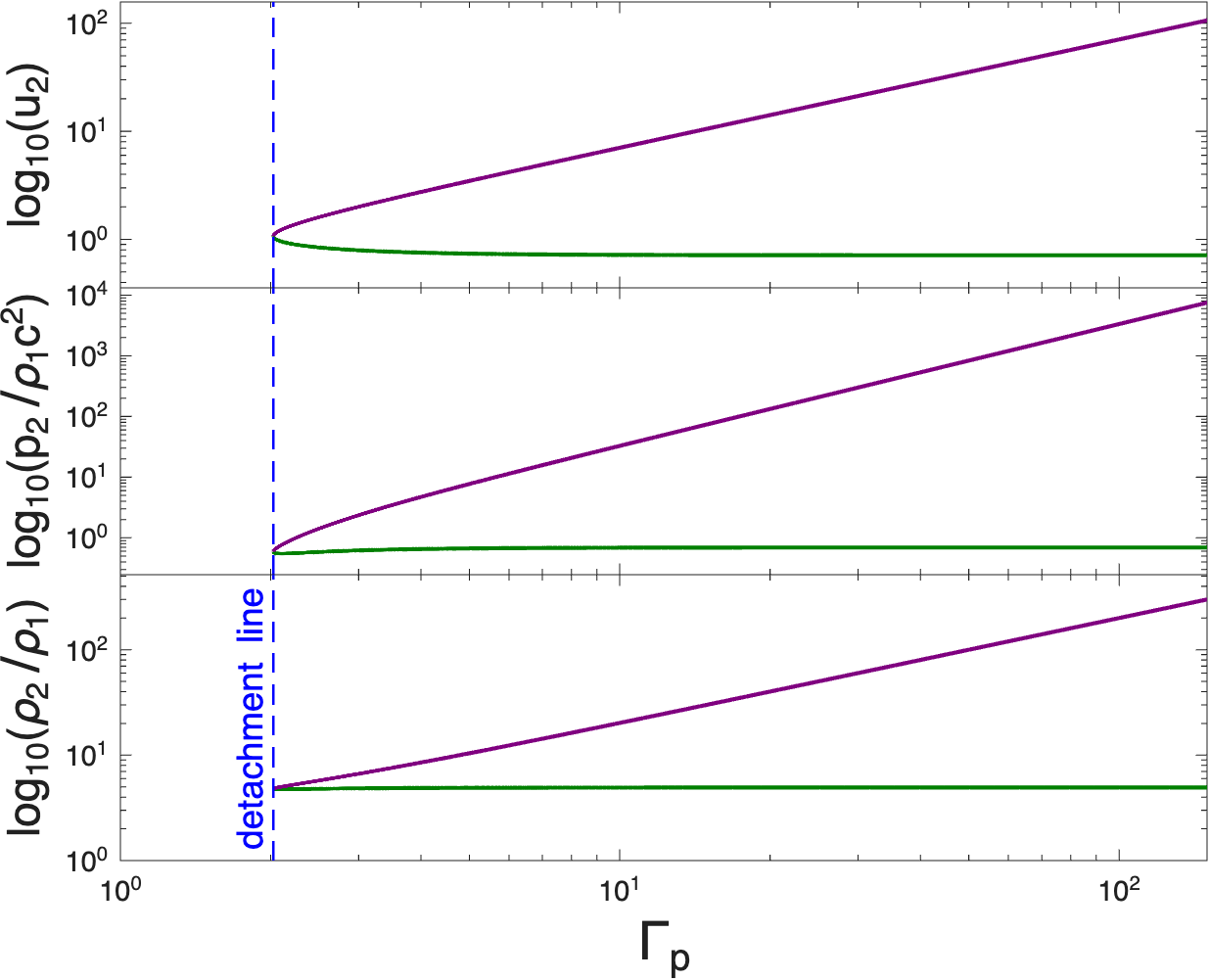}
\vspace{-0.5cm}
\caption{The same hydrodynamic variables in region 2 as in Fig.~\ref{fig:SW_solutions_a1}, with the same color-coding, but as a function
of $\Gamma_p$ (the Lorentz factor corresponding to the boost between the lab frame $S$
and the steady-state frame $S'$) in the sub-luminal attachment region between the
detachment and luminal lines.}
\label{fig:SW_solutions_Gp}
\end{figure}

Figure~\ref{fig:SW_solutions_Gp} shows the same hydrodynamic variables as Fig.~\ref{fig:SW_solutions_a1} in the shocked region 2, but as a function of the Lorentz factor $\Gamma_p$ corresponding to a boost between the lab frame $S$ and the steady-state frame $S'$. 
The divergence of $u_2$, $p_2$, and $\rho_2$ toward the luminal line (i.e. at large $\Gamma_p$ values) for the strong shock solution can clearly be seen
here. Moreover, it can be seen that in this limit, $u_2$, $\rho_2\propto\Gamma_p$, while
$p_2\propto\Gamma_p^2$. 

This can be understood as follows. As $\Gamma_p\to\infty$ for a fixed $u_1$, $\tan\alpha_1\to u_1$ approaches a constant value, but in frame $S'$ we have $\tan\alpha'_1=\Gamma_p^{\,-1}\tan\alpha_1\to0$. For $\Gamma_p\gg1$ this implies $1\gg\alpha'_1\approx\tan\alpha'_1\approx u_1/\Gamma_p\propto\Gamma_p^{\,-1}$. Similarly, $\Gamma'_1=\Gamma_p\Gamma_1(1-\beta_p\beta_1\sin\alpha_1)\to\Gamma_p/\Gamma_1\propto\Gamma_p$. In the steady state frame $S'$ the velocity of the unshocked region 1 is nearly parallel to the wall ($\alpha'_1\ll1$), such that for the strong shock solution $\alpha'_2\approx\pi/2$ and the shock must be almost perpendicular to achieve the required small deflection to make the velocity of the fluid in region 2 exactly parallel to the wall. Such a nearly perpendicular relativistic shock (with $1\ll\Gamma'_1\propto\Gamma_p$) slows down the fluid at region 2 to a mildly 
relativistic velocity in frame $S'$ corresponding to $u_2\propto\Gamma_p$ in 
frame $S$. Similarly, its compression ratio scales as $\Gamma'_1\propto\Gamma_p$ explaining why $\rho_2\propto\Gamma_p$, while 
the downstream pressure scales as the upstream ram pressure, $p_2\propto (u'_1)^2\approx(\Gamma'_1)^2\propto\Gamma_p^2$.
This behavior is again similar that for relativistic shock reflection \citep{Granot-Rabinovich-24,Bera+24}.

\subsection{The Solution Along the Sonic\,/\,Detachment Line}
\label{sec:sonic}

\begin{figure}
\centering
\includegraphics[trim={0cm 0cm 0cm 0cm},clip,scale=0.6,width=0.48\textwidth]{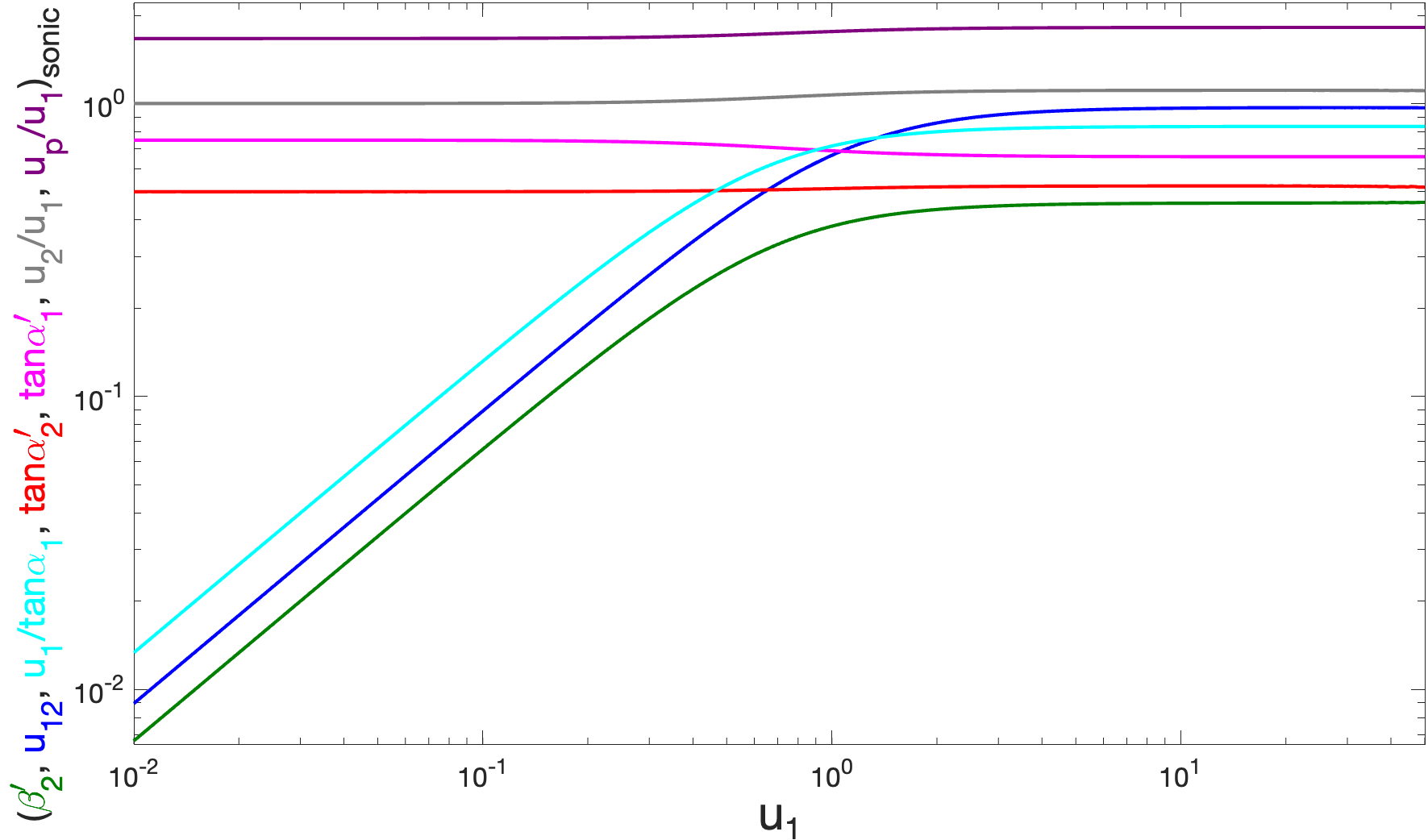}
\vspace{-0.5cm}
\caption{The properties of the flow along the sonic line.}
\label{fig:sonic-line}
\end{figure}

Figure~\ref{fig:sonic-line} illustrates the properties of the flow along the \textbf{sonic line}, which for an initially cold shell coincides with the \textbf{detachment line}. The sonic line must always be in the sub-luminal region and can therefore conveniently be analyzed in the rest frame $S'$ of point $P$, where the flow is steady. In this frame the oblique shock jump conditions and the sonic condition read
\begin{eqnarray}
\rho_1u'_1\sin\alpha'_+&=&\rho_2u'_2\sin\alpha'_2\ ,
\\ \nonumber
\rho_1c^2u^{\prime\,2}_1\sin^2\!\alpha'_+&=&w_2u^{\prime\,2}_2\sin^2\!\alpha'_2+p_2\ ,
\\ \label{eq:Sp0}
\rho_1c^2\Gamma'_1u'_1\sin\alpha'_+&=&w_2\Gamma'_2u'_2\sin\alpha'_2\ ,
\\ \nonumber
\beta'_1\cos\alpha'_+&=&\beta'_2\cos\alpha'_2\ ,
\\ \nonumber
(\beta'_2)^2 &=& c_{s,2}^2 = \frac{3\Theta_2^2+5\Theta_2\sqrt{\Theta_2^2+\frac{4}{9}}}{12\Theta_2^2+2+12\Theta_2\sqrt{\Theta_2^2+\frac{4}{9}}}\ ,
\end{eqnarray}
denoting $\alpha'_+=\alpha'_1+\alpha'_2$ and $\Theta_2=p_2/\rho_2c^2$.

The flow along the sonic line has particularly simple limits in the relativistic or Newtonian limits. 
In the \textbf{Newtonian limit} ($u_1\ll1$) the adiabatic index is $\hat{\gamma}_2=\frac{5}{3}$ and Eqs.~(\ref{eq:Sp0}) reduce to
\begin{eqnarray}\nonumber
\rho_1v'_1\sin\alpha'_+&=&\rho_2v'_2\sin\alpha'_2\ ,
\\ \nonumber
\rho_1v^{\prime\,2}_1\sin^2\!\alpha'_+&=&\rho_2v^{\prime\,2}_2\sin^2\!\alpha'_2+p_2\ ,
\\
v^{\prime\,2}_1\sin^2\alpha'_+&=&5\frac{p_2}{\rho_2}+v^{\prime\,2}_2\sin^2\alpha'_2\ ,
\\ \nonumber
v'_1\cos\alpha'_+&=&v'_2\cos\alpha'_2\ ,
\\ \nonumber
v^{\prime\,2}_2 &=& \frac{5}{3}\frac{p_2}{\rho_2}\ ,
\end{eqnarray}
and the flow approaches an asymptotic state where the limiting values are given by:
\begin{eqnarray}\nonumber
\frac{\rho_2}{\rho_1}&\to&4\ ,\quad\quad\quad\quad\quad\quad\ \ \ \,
\frac{p_2}{\rho_1v_1^2}\to\frac{16}{15}\ ,
\\ \nonumber
\frac{u_2}{u_1} &\to& \frac{\beta_2}{\beta_1} \to 1\ ,\quad\quad\quad\quad\ 
\frac{u'_1}{u'_2} \to \frac{\beta'_1}{\beta'_2} \to 2\ ,
\\  \nonumber
\frac{u_{12}}{u_1} &\to&  \frac{u_{12}}{u_2}\to\frac{\beta_{12}}{\beta_1} \to \frac{\beta_{12}}{\beta_2} \to \frac{2}{\sqrt{5}}\ ,
\\  
\label{eq:sonic-Newt}
\frac{u_p}{u_1} &\to& \frac{u_p}{u_2}\to\frac{\beta_p}{\beta_1}\to \frac{\beta_p}{\beta_2} \to \frac{5}{3}\ ,
\\  \nonumber
\tan\alpha'_1 &\to& \tan\alpha_1 \to \frac{3}{4}\ ,
\quad\quad\quad
\tan\alpha'_2 \to \tan\alpha_2 \to \frac{1}{2}\ ,
\\  \nonumber
\sin\alpha'_1 &\to& \sin\alpha_1 \to \frac{3}{5}\ ,
\quad\quad\quad
\sin\alpha'_2 \to \sin\alpha_2 \to \frac{1}{\sqrt{5}}\ ,
\\  \nonumber
\cos\alpha'_1 &\to& \cos\alpha_1 \to \frac{4}{5}\ ,
\quad\quad\ \ \ \,
\cos\alpha'_2 \to \cos\alpha_2 \to \frac{2}{\sqrt{5}}\ .
\end{eqnarray}

In the \textbf{relativistic limit} ($\Gamma_1\approx u_1\gg1$) 
we have $\alpha_1,\,\alpha_2\approx\frac{\pi}{2}$ along the sonic line, so it is convenient to use the angle $\bar{\alpha}_1=\frac{\pi}{2}-\alpha_1$ 
to express the solution to the above equations in terms of $a=\Gamma_1\bar{\alpha}_1$ and $b=\Gamma_2/\Gamma_1$. Taking the appropriate limit of Eqs.~(\ref{eq:Sp0}) and using the Lorentz transformation of the angles ($\tan\alpha_{i} = \Gamma_p\tan\alpha'_{i}$ for $i=1,\,2$) leads to the following limiting values:
\begin{eqnarray}\nonumber
\beta'_1 &=& \frac{u_1}{\tan\alpha_1} \to \cos\alpha'_1\to a = 0.835305602158\ ,
\\ \nonumber
\frac{u_2}{u_1} &\to& \frac{\Gamma_2}{\Gamma_1}\to b = 1.109793088952\ ,
\\ \nonumber
\Gamma'_1 &=& \frac{\Gamma_p}{\Gamma_1}\to\frac{u_p}{u_1} \to \frac{1}{\sqrt{1-a^2}}\approx1.818889816772\ ,
\\  \nonumber
\tan\alpha'_1 &\to&
\frac{1}{u'_1} \to \frac{\sqrt{1-a^2}}{a}\approx0.658185354403\ ,
\\ \nonumber
\beta'_2 &=& c_{s,2}\to\frac{1-b^2+a^2b^2}{1+b^2-a^2b^2}\approx 0.457427107756\ ,
\\ 
\Gamma'_2 &\to& \frac{1+b^2-a^2b^2}{2b\sqrt{1-a^2}}\approx 1.124546795925\ ,
\\ \nonumber
\Gamma_{12}&\to& \frac{1+b^2+a^2b^2}{2b}\approx1.392601967012\ ,
\\ \nonumber
u_{12} &\to& 0.969195665759\ ,
\\  \nonumber
\tan\alpha'_2 &\to&
\frac{\sqrt{1-a^2}(b^2+a^2b^2-1)}{a(b^2-a^2b^2+1)}\approx0.523274827366\ ,
\\ \nonumber
\frac{\rho_2}{\rho_1} &=& 4\Gamma_{12}\approx5.570407868047\ ,
\\ \nonumber
\frac{p_2}{\rho_1c^2}&=&\frac{4}{3}u_{12}^2\approx1.252453651368\ .
\end{eqnarray}

\section{Steady-state analysis in the sub-luminal case}
\label{sec:cons-steady}

In the sub-luminal case ($v_p<c$) there is a reference frame $S'$ where point $P$ is at rest and the flow is steady. Therefore, a ruler at rest in $S'$ oriented parallel to the wall will Lorentz contract in frame $S$, $L_\parallel=L'_\parallel/\Gamma_p$, while in the perpendicular direction the length remains unchanged, $L_\perp=L'_\perp$. Hence, the angle of the shock front relative to the wall transforms as 
\begin{equation}
\tan\alpha_{i} = \frac{L_{i,\perp}}{L_{i,\parallel}} = \Gamma_p\frac{L'_{i,\perp}}{L'_{i,\parallel}} = \Gamma_p\tan\alpha'_{i}\ ,\quad(i=1,\,2)\ .
\end{equation}
Note that the relative Lorentz factor of regions 1 and 2 can be expressed in this frame as
\begin{equation}
\Gamma_{12} = \Gamma'_1\Gamma'_2(1-\beta'_1\beta'_2\cos\alpha'_1)\;.
\label{eq:Gamma12p}
\end{equation}

As can be seen in the bottom panel of Fig.~\ref{fig:shell-wall}, in frame $S'$ the upstream velocity $\textbf{\textit{v}}'_1$ is parallel to he vacuum interface and the velocity deflection angle $\chi'$ is simply $\alpha'_1$. Therefore, conveniently denoting  $\alpha'_+=\alpha'_1+\alpha'_2$,
the oblique shock jump conditions in this frame can be written as
\begin{eqnarray}\nonumber
\rho_1u'_1\sin\alpha'_+ &=& \rho_2u'_2\sin\alpha'_2\ ,
\\ \nonumber
w_1(u'_1\sin\alpha'_+)^2+p_1 &=& w_2(u'_2\sin\alpha'_2)^2+p_2\ ,
\\ \label{eq:Sprime0}
w_1\Gamma'_1u'_1\sin\alpha'_+ &=& w_2\Gamma'_2u'_2\sin\alpha'_2\ ,
\\ \nonumber
\beta'_1\cos\alpha'_+ &=& \beta'_2\cos\alpha'_2\ .
\end{eqnarray}
In accordance with the usual convention used in the context of Newtonian flow, we represent the angle of the shock relative to the upstream flow as $\phi'$, such that $\phi'=\alpha_+'$. Similarly, we denote the deflection angle as $\chi'$, indicating that $\chi'=\alpha_1'$ and $\alpha'_2=\phi'-\chi'$. Thus we can express Eq.~(\ref{eq:Sprime0}) as
\begin{eqnarray}\nonumber
\rho_1u'_1\sin\phi' &=& \rho_2u'_2\sin(\phi'-\chi')\ ,
\\ \nonumber
w_1(u'_1\sin\phi')^2+p_1 &=& w_2[u'_2\sin(\phi'-\chi')]^2+p_2\ ,
\\ \label{eq:Sprime}
w_1\Gamma'_1u'_1\sin\phi' &=& w_2\Gamma'_2u'_2\sin(\phi'-\chi')\ ,
\\ \nonumber
\beta'_1\cos\phi' &=& \beta'_2\cos(\phi'-\chi')\ .
\end{eqnarray}

%
Specifying to a cold shell, $p_1=0$ and $w_1=\rho_1c^2$, these equations reduce to
\begin{eqnarray}\nonumber
\rho_1u'_1\sin\phi' &=& \rho_2u'_2\sin(\phi'-\chi')\ ,
\\ \nonumber
\rho_1c^2(u'_1\sin\phi')^2 &=& w_2[u'_2\sin(\phi'-\chi')]^2+p_2\ ,
\\
\rho_1c^2\Gamma'_1u'_1\sin\phi' &=& w_2\Gamma'_2u'_2\sin(\phi'-\chi')\ , \label{eq:Sprime2}
\\ \nonumber
\beta'_1\cos\phi' &=& \beta'_2\cos(\phi'-\chi')\ .
\end{eqnarray}

\smallskip
\noindent\textbf{Deflection angle}:
Combining the first and fourth relations in Eqs.~(\ref{eq:Sprime2}) with Eq.~(\ref{eq:1Dshock}), and in particular $\frac{\rho_2}{4\rho_1}=\Gamma_{12}=\Gamma'_1\Gamma'_2(1-\beta'_1\beta'_2\cos\chi')$, yields an explicit expression relating the deflection angle $\chi'$ to the shock angle $\phi'$ for a given upstream flow velocity 
\begin{equation}
\beta_1^{\prime\,2} = \frac{\cos(\phi'-\chi')[4\cos\phi'\sin(\phi'-\chi') - \sin\phi'\cos(\phi'-\chi')]}{\cos^2\!\phi'[4\cos\chi'\sin(\phi'-\chi') - \sin\phi']} \,.
\end{equation}

\begin{figure}[t]
\centering
\includegraphics[trim={0cm 0cm 0.05cm 0cm},clip,scale=1,width=0.48\textwidth,height=0.33\textwidth]{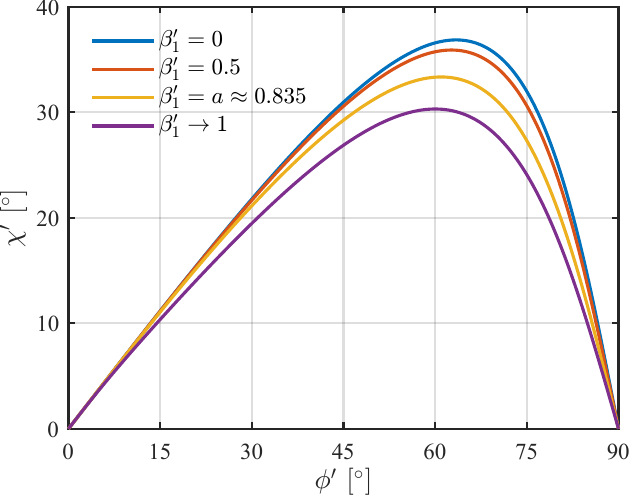}
\caption{The deflection angle $\chi'$  as a function of the shock angle $\phi'$ (between the upstream velocity and the shock front).}
\label{fig:deflection}
\end{figure}

Figure~\ref{fig:deflection} shows the dependence of $\chi'$ on $\phi'$ for several representative values of $\beta'_1$. In the Newtonian limit ($\beta'_1 \to 0$), this relation neatly reduces to the well-known classical expression:
\begin{equation}\label{eq:chi-phi-Sp}
\tan\chi' = \frac{3\tan\phi'}{4+\tan^2\!\phi'} \, .
\end{equation}
This formula describes an oblique shock in a cold gas where the upstream sound speed vanishes ($c_s \to 0$), corresponding to an infinite upstream classical Mach number, $\mathcal{M}'_1 \equiv v'_1/c_s \to \infty$. \citep{Landau+Lifshitz1987} Maximizing this expression readily yields $\tan\chi'_{\rm max} = 3/4$ at a shock angle of $\tan\phi' = 2$. This perfectly aligns with Eq.~(\ref{eq:sonic-Newt}), reflecting the fundamental classical result that the sonic and detachment lines coincide in the limit of infinite Mach number.


\smallskip
\noindent\textbf{Downstream speed}:
Combining the ratio of the third and first relations in Eqs.~(\ref{eq:Sprime2}), $\Gamma'_1=\Gamma'_2w_2/(\rho_2c^2)$, with Eq.~(\ref{eq:1Dshock}), and in particular $\frac{\rho_2}{4\rho_1}=\Gamma_{12}=\Gamma'_1\Gamma'_2(1-\beta'_1\beta'_2\cos\chi')$ and $w_2 = 4\Gamma_{12}^2\left(1+\frac{1}{3}\beta_{12}^2\right)\rho_1c^2$, yields a cubic equation for $\beta_2'$:
\begin{eqnarray}\nonumber
    3\beta_1'\cos\alpha_1'\beta_2^{\prime\,3}+[(1-4\cos^2\alpha'_1)\beta_1^{\prime\,2} - 4]\beta_2^{\prime\,2} &
    \\ \label{eq:beta2p}
    +5\beta_1'\cos\alpha_1'\beta'_2-\beta_1^{\prime\,2}&=0\;.
\end{eqnarray}

\smallskip
\noindent\textbf{Detachment angle}:
At the detachment point, the weak and strong shock solutions merge into a single repeated root, which dictates that the discriminant of this cubic equation must vanish. Equating the discriminant to zero yields (upon dividing by 4) the following polynomial constraint:
\begin{equation}\label{eq:alpha_det_p}
\begin{aligned}
    (-64\beta_1^{\prime\,8} + 100\beta_1^{\prime\,6})\cos^6\!\alpha'_{\rm det} &\\
    + \left(48\beta_1^{\prime\,4} + 28\beta_1^{\prime\,2} - 175\right)\beta_1^{\prime\,4}\cos^4\!\alpha'_{\rm det} &\\
    + (-12\beta_1^{\prime\,8} - 26\beta_1^{\prime\,6} + 28\beta_1^{\prime\,4} + 100\beta_1^{\prime\,2})\cos^2\!\alpha'_{\rm det} &\\
    + \beta_1^{\prime\,2}(\beta_1^{\prime\,2} - 4)^3 &= 0 \;.
\end{aligned}
\end{equation}
This expression represents a cubic equation in terms of $\cos^2\alpha'_{\rm det}$, which admits an analytic solution as a function of the upstream velocity $\beta_1'$. Within the physically relevant range $0 < \beta_1' < 1$, exactly one of its three roots is real and bounded by the interval $(0,1)$. 

\begin{figure}[h]
\centering
\includegraphics[trim={0cm 0cm 0.05cm 0cm},clip,scale=1,width=0.48\textwidth,height=0.33\textwidth]{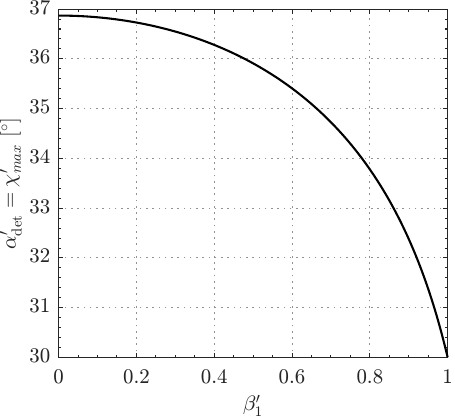}
\vspace{-0.4cm}
\caption{The detachment angle $\alpha'_{\rm det}=\chi'_{\rm max}$ as a function of the upstream velocity $\beta'_1$.}
\label{fig:max_deflection}
\end{figure}

To explore the physical boundaries of the detachment condition, we examine the roots of Eq.~(\ref{eq:alpha_det_p}) at the two asymptotic velocity limits. In the Newtonian regime ($\beta_1' \ll 1$), keeping only the leading-order terms of $O(\beta_1'^2)$ (and dividing by 16) simplifies the condition to:\begin{equation}
25\cos^2\!\alpha'_{\rm det} - 16 = 0\;,
\end{equation}
which yields $\cos\alpha'_{\rm det} = 4/5$, corresponding to the classical detachment angle of $\alpha'_{\rm det} \approx 36.87^\circ$. This recovers the well-known infinite Mach number limit for a cold gas. Conversely, in the ultra-relativistic limit ($\beta_1' \to 1$), the detachment condition reduces (upon division by 36) to the following cubic polynomial in terms of $\cos^2\alpha'_{\rm det}$:
\begin{equation}
\begin{aligned}\nonumber
4\cos^6\!\alpha'_{\rm det} - 11\cos^4\!\alpha'_{\rm det} + 10\cos^2\!\alpha'_{\rm det} - 3 &= \\(\cos^2\!\alpha'_{\rm det}-1)^2(4\cos^2\!\alpha'_{\rm det}-3) &= \,0\;.
\end{aligned}
\end{equation}
Factoring out the trivial solutions ($\cos^2\!\alpha'_{\rm det}=1$), the physically relevant root is $\cos^2\!\alpha'_{\rm det} = 3/4$, which implies an exact maximum deflection angle of $\alpha'_{\rm det} = 30^\circ$. This reduction in the detachment angle compared to the Newtonian case reflects the increased effective inertia of the relativistically hot downstream fluid. The coupling of pressure and internal energy to the fluid's inertia ``stiffens'' the flow, dynamically restricting its ability to undergo sharp deflections while remaining attached to the wall. Figure~\ref{fig:max_deflection} illustrates the dependence of the detachment angle $\alpha'_{\rm det}$ on the upstream velocity $\beta'_1$.

\smallskip 
\noindent\textbf{Proof for the equality of the sonic and detachment lines}:
The sound speed in the shocked region 2 can be determined by substituting $\Theta_2 = \frac{\Gamma_{12}^{\,2} - 1}{3\Gamma_{12}}$ (Eq. (\ref{eq:1Dshock})) into Eq. (\ref{eq:Taub_soundspeed}), yielding:
\begin{equation}
    \beta_{c_s}^2  = \frac{(4\Gamma_{12}^{\,2} + 1)(\Gamma_{12}^{\,2} - 1)}{3\Gamma_{12}^{\,2}(4\Gamma_{12}^{\,2} - 1)}.
    \label{eq:soundspeed}
\end{equation}
To prove that the sonic line and the detachment line perfectly coincide, we must show that the downstream fluid velocity equals the local sound speed exactly at the detachment limit.

Dividing the energy conservation equation by the mass conservation equation in Eqs.~(\ref{eq:Sprime2}), and applying the strong shock enthalpy condition $w_2 = 4\Gamma_{12}^{\,2}\left(1+\frac{1}{3}\beta_{12}^2\right)\rho_1c^2$ from Eqs.~(\ref{eq:1Dshock}), we obtain:
\begin{equation}
    \frac{\Gamma'_1}{\Gamma'_2} = \frac{4\Gamma_{12}^{\,2} - 1}{3\Gamma_{12}}\;.
    \label{eq:Gamma1}
\end{equation}
By utilizing the general expression for the relative Lorentz factor $\Gamma_{12}$ (Eq.~(\ref{eq:Gamma12p})) alongside Eq.~(\ref{eq:Gamma1}), we can express the deflection angle $\chi'=\alpha'_1$ purely in terms of Lorentz factors:
\begin{equation}
    \cos\chi' = \frac{\Gamma_{12}(3\Gamma_1^{\prime\,2} - 4\Gamma_{12}^{\,2} + 1)}{\sqrt{\Gamma_1^{\prime\,2} - 1}\sqrt{9\Gamma_1^{\prime\,2}\, \Gamma_{12}^{\,2} - (4\Gamma_{12}^{\,2} - 1)^2}}\;.
\end{equation}
The detachment condition corresponds to the maximum possible physical deflection angle, meaning we maximize $\chi'$ (or equivalently, minimize $\cos^2\!\chi'$) for a fixed $\Gamma'_1$. Setting the derivative of $\cos^2\!\chi'$ with respect to $\Gamma_{12}^{\,2}$ to zero yields the exact detachment criterion:
\begin{equation}\label{eq:detachment_condition}
\Gamma_1^{\prime\,2} = \frac{(4\Gamma_{12}^{\,2} - 1)^3}{3(8\Gamma_{12}^{\,4} + 1)}\;.
\end{equation}
Finally, substituting this criterion back into the kinematic relation $\beta_2^{\prime\,2} = 1 - (\Gamma'_2)^{-2}$ via Eq.~(\ref{eq:Gamma1}), we find that at the detachment point, the downstream velocity squared is:
\begin{equation}
    \beta_2^{\prime\,2} = \frac{(\Gamma_{12}^{\,2} - 1)(4\Gamma_{12}^{\,2} + 1)}{3\Gamma_{12}^{\,2} (4\Gamma_{12}^{\,2} - 1)}.
\end{equation}
This result is identically equal to Eq.~(\ref{eq:soundspeed}), thereby providing a rigorous proof that the sonic line and the detachment line coincide.

From a physical perspective, this exact coincidence arises because the incident shell is cold, meaning the initial pressure in region 1 vanishes. In the attachment region, the stable weak-shock solution ensures that the downstream flow remains supersonic relative to the collision point. Consequently, acoustic signals are swept downstream and cannot propagate back to the shock front and in particular cannot affect the point $P$ where the shock attaches to the wall. However, as the deflection angle increases, the downstream velocity steadily decreases. Precisely at the detachment threshold, the downstream fluid transitions to a subsonic state. This critical transition establishes causal acoustic contact: pressure waves can now travel upstream, communicating the downstream pressure build-up to the shock front. This acoustic feedback physically pushes the shock away from the wall, forcing it to detach and allowing the high-pressure fluid to spill into the vacuum.

\begin{figure}[h!]
\includegraphics[trim={0.0cm 2.58cm 0.54cm 1.40cm},clip,scale=1,width=0.30\textwidth]{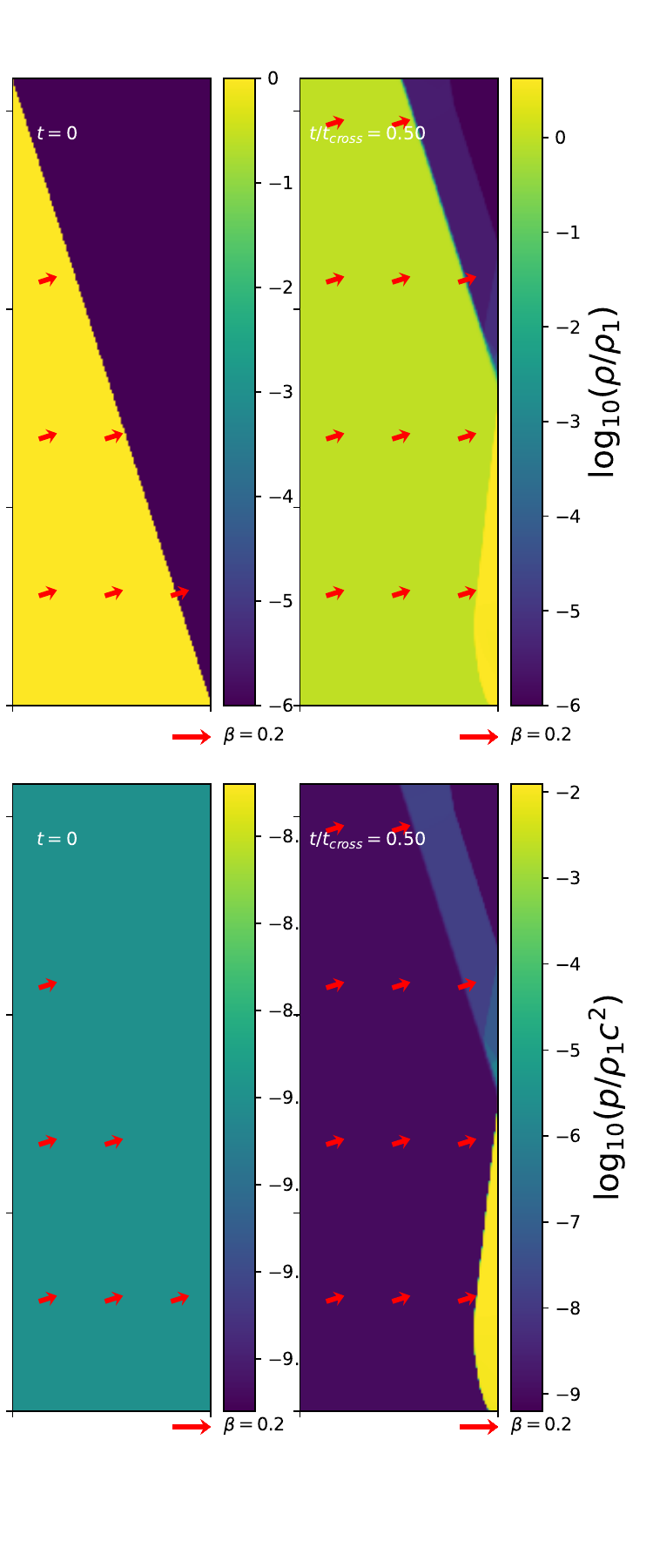}
\caption{Snapshot from our relativistic hydrodynamic numerical simulations of a cold shell moving at a proper speed $u_1=0.1$ in the lab frame $S$ and colliding with a perfect reflecting wall at an incidence angle $\alpha_1=0.306$. We show the normalized proper rest-mass density ($\rho/\rho_1$; \textit{top panels}) and pressure ($p/\rho_1c^2$; \textit{bottom panels}) at the start of our simulation (\textit{left panels}) and at a later time (\textit{right panels}) when point $P$ has crossed half of the computational domain.}	
\label{fig:sim-frame-S-attached}
\end{figure}

\section{Numerical Simulations}
\label{sec:simulations}

The numerical methods used here largely follow those described in detail in our previous work\cite{Bera+24}, where we use the \textsc{pluto} code \citep{Mignone+2007} to solve the special relativistic hydrodynamic equations in a fixed linear spaced grid.
The setup is also similar but adjusted to the current problem of a cold shell propagating in vacuum and obliquely colliding with an ideal reflecting wall (see Fig.~\ref{fig:shell-wall}). Simulations are performed in either the lab frame $S$ where the the velocity of the cold shell $\bm{v}_1$ is perpendicular to its vacuum interface, and in the steady-state frame $S'$ (which exists only in the sub-luminal region) where $\bm{v}'_1$ is parallel to the vacuum interface.

\begin{figure}[h!]
\includegraphics[trim={2.26cm 2.58cm 0.54cm 1.40cm},clip,scale=1,width=0.28\textwidth]{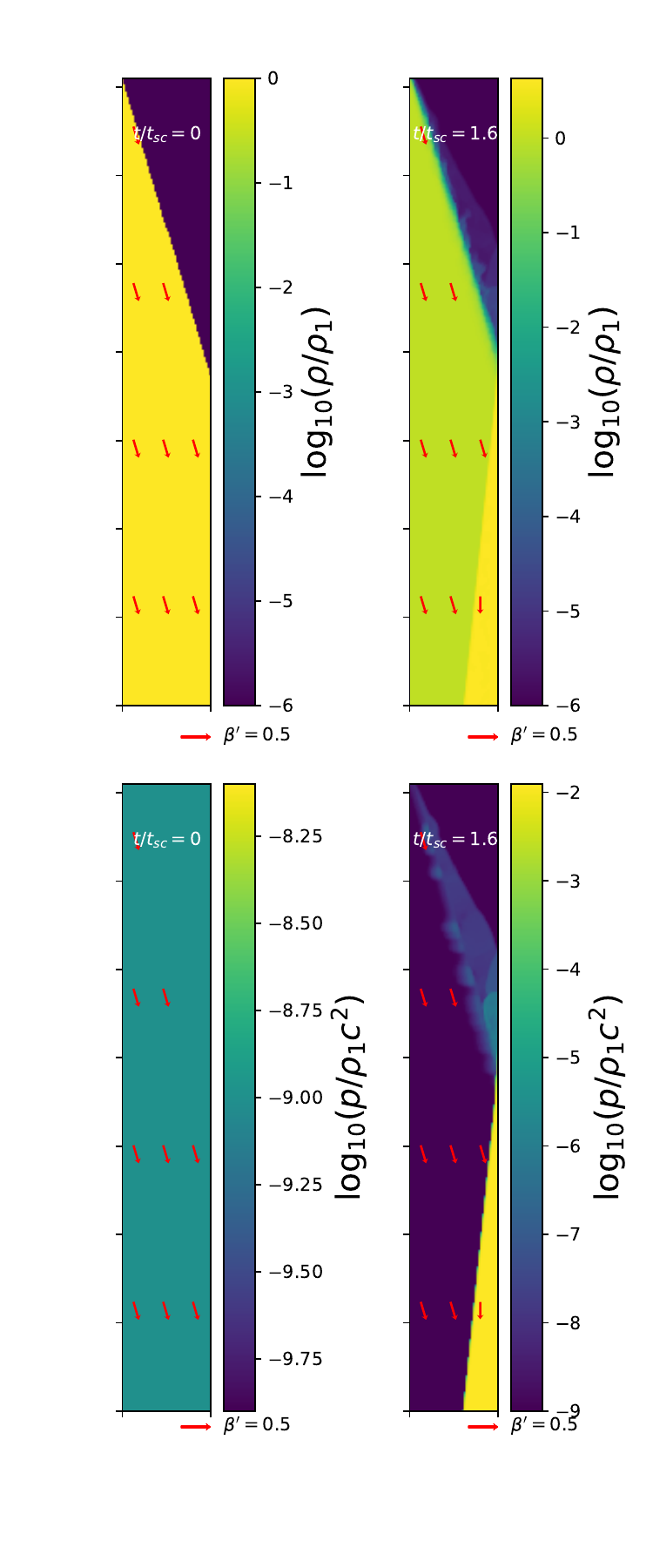}
\caption{Snapshots from our relativistic hydrodynamic numerical simulations of a cold shell moving at a proper speed $u_1=0.1$ in the lab frame $S$ and colliding with a perfect reflecting wall at an incidence angle $\alpha_1=0.306$. The simulation shown here is performed in the steady-state frame $S'$, which moves relative to frame $S$ along the wall at a normalized velocity $\beta_p=\frac{\beta_1}{\sin\alpha_1}=0.330$. The snapshot format is similar to Fig.~\ref{fig:sim-frame-S-attached} where at the second time here (\textit{right panels}) a steady-state has been reached.}	
\label{fig:sim-frame-Sp-attached}
\end{figure}

\begin{figure}
\includegraphics[trim={0.18cm 1.55cm 3.95cm 0.90cm},clip,scale=1,width=0.99\columnwidth]{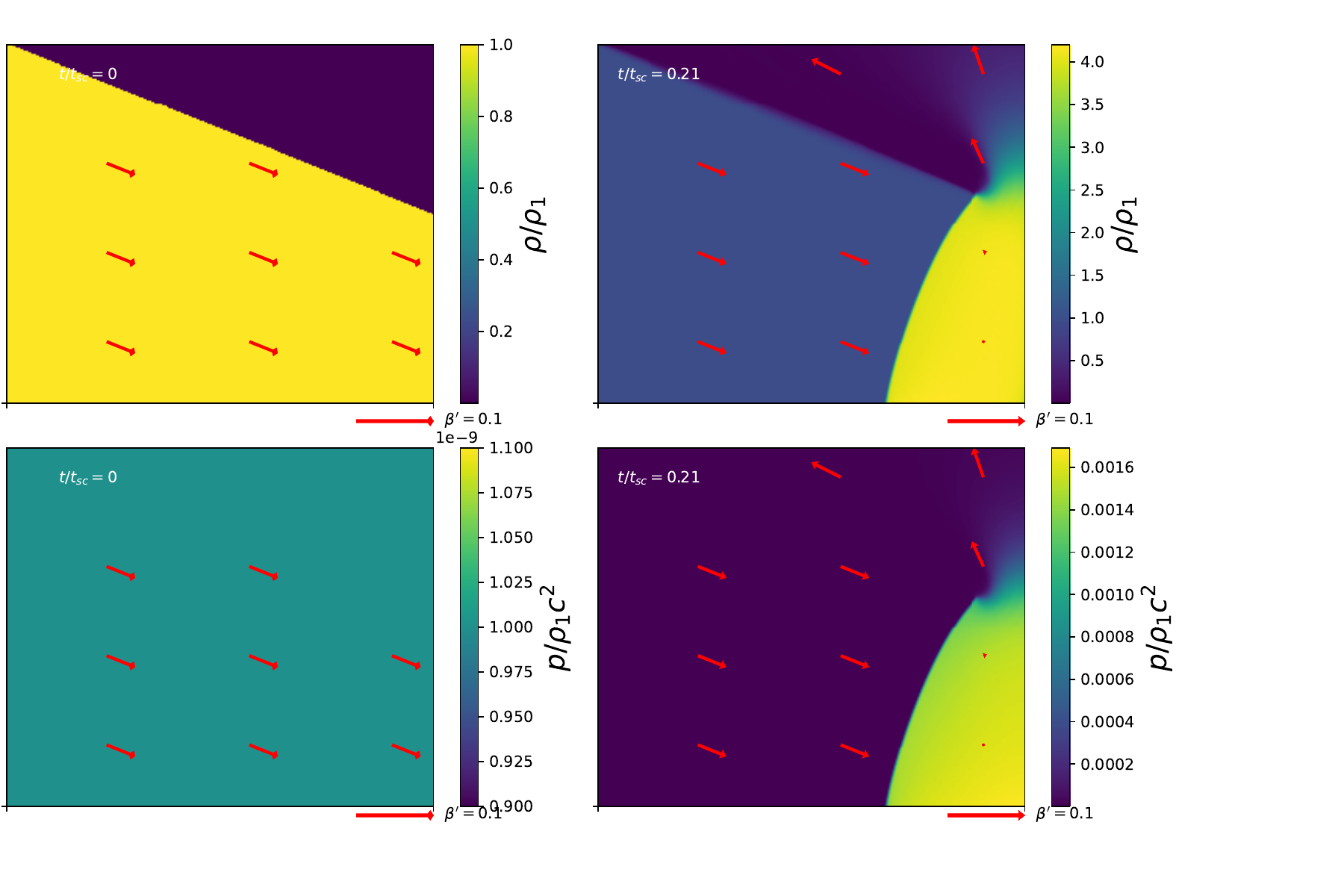}
\caption{Snapshots from a simulation preformed in frame $S'$,
with $u_1=0.1$ similar to Fig.~\ref{fig:sim-frame-Sp-attached},  but $\alpha_1=1.19>\alpha_{\rm det}$ placing it in the detachment region and implying $\beta_p=\frac{\beta_1}{\sin\alpha_1}=0.107$. As expected, point $P$ detaches from the wall and shocked fluid spills from region 2 into the vacuum region 0.}	
\label{fig:sim-frame-Sp-detached}
\end{figure}

Figure~\ref{fig:sim-frame-S-attached} shows snapshots of the density and pressure from a simulation that we have performed in the lab frame $S$. The two left panels show our initial conditions, where $\bm{v}_1$ normal to the vacuum interface and the pressure is set to a finite but very low floor value (of $p_1=10^{-9}\rho_1c^2$) for numerical stability reasons. The proper speed of the incident cold shell is $u_1=0.1$ and it collides with the perfect reflecting wall at an incidence angle $\alpha_1=0.306$. The interaction point $P$ moves along the wall at a velocity $v_p=\frac{v_1}{\sin\alpha_1}\approx0.330c$. A shock forms that makes an angle $\alpha_2$ with the wall and bounds a shocked region 2. The lower boundary condition has some effect on the lower part of the computational box, and this part is therefore excluded when extracting the properties of region 2. The parameters used in this run ($u_1=0.1$, $\alpha_1=0.306$) place it well within the sub-luminal attachment region ($\alpha_{\rm lum}<\alpha_1<\alpha_{\rm det}$) where $\alpha_{\rm lum}=\arctan(u_1)\approx0.0997$ and $\alpha_{\rm det}\approx0.649$), and indeed point $P$ remains attached to the wall.

Figure~\ref{fig:sim-frame-Sp-attached} shows snapshots of the density and pressure from a simulation that we have performed in the steady-state frame $S'$, for the same $S$-frame parameter values ($u_1=0.1$, $\alpha_1=0.306$). The two left panels show our initial conditions, where $\bm{v}'_1$ is parallel to the vacuum interface and the pressure is set to the same finite but very low floor value (of $p_1=10^{-9}\rho_1c^2$). 
On a timescale of the order of the lateral sound crossing time of the shocked region 2 a steady state is established, which is depicted in the right panels, and shown below to correspond to the weak shock solution. We find that this solution is naturally chosen by the system and is stable. There is a very small amount of mass  that leaks into the vacuum region 0, both through the vacuum interface (via a weak rarefaction wave driven by the small but finite pressure in region 1) and through the interaction point $P$ (because of the finite numerical resolution near this point). 

\begin{figure}
\includegraphics[trim={0.55cm 1.45cm 1.18cm 2.43cm},clip,scale=1,width=.99\columnwidth,height=1.1\columnwidth]{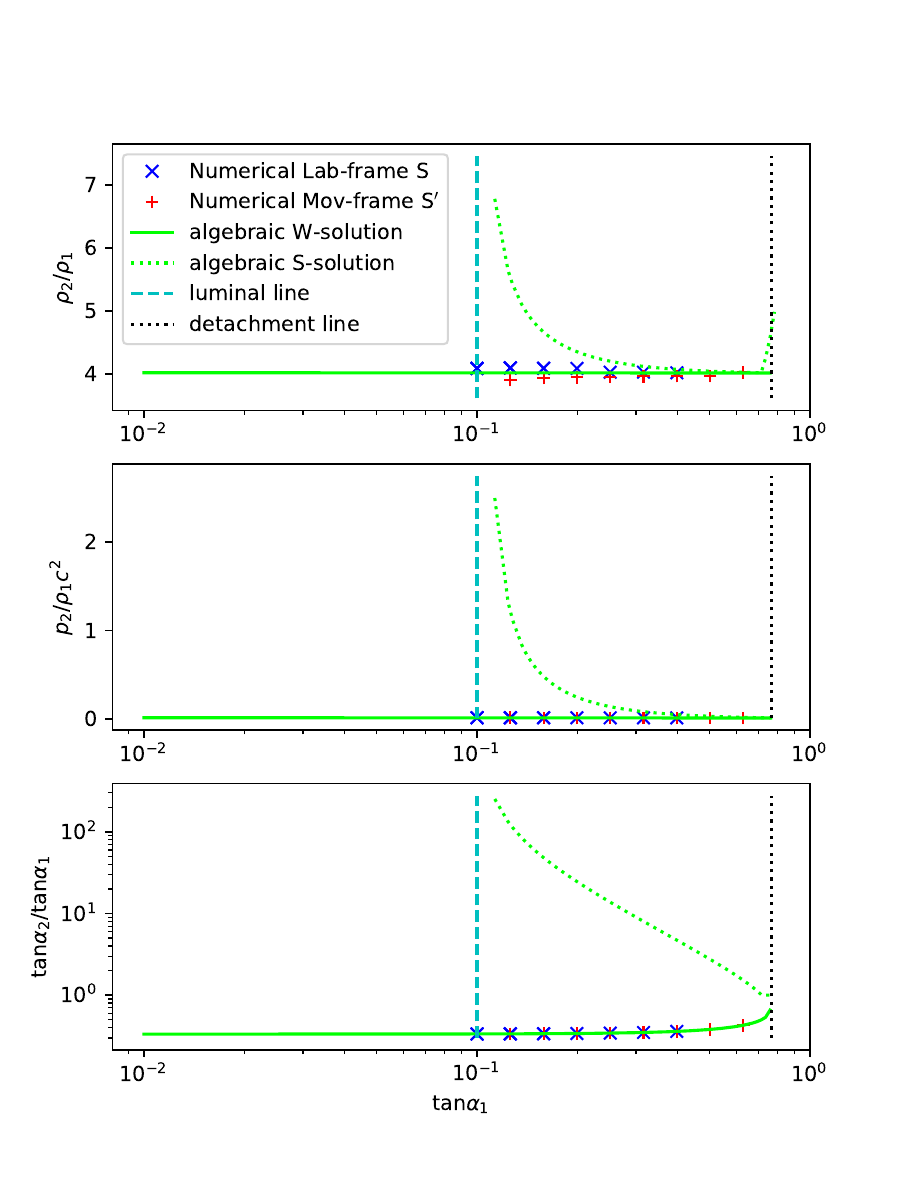}
\vspace{-0.1cm}
\caption{Comparison of our analytic weak-shock (solid \green{green} lines) and strong-shock (dotted \green{green} lines) solutions for 
$u_1=0.1$ to the results of our relativistic hydrodynamic numerical simulations in the lab frame $S$ (blue \blue{X}'s) and in the steady-state frame $S'$ (red \red{+}'s). Results are shown for the shocked region 2: normalized proper density ($\rho_2/\rho_1$; \textit{top panel}), normalized pressure ($p_2/\rho_1c^2$; \textit{middle panel}) and $\tan\alpha_2/\tan\alpha_1$ (\textit{bottom panel}).
Our chosen $u_1=0.1$ intersects the luminal and detachment lines at $\alpha_{\rm lum}=\arctan(u_1)\approx0.0996686525$ (vertical dashed \cyan{cyan} line) and $\alpha_{\rm det}\approx0.6490538467$ (vertical dotted black line), respectively. The weak shock solution is naturally chosen by the system, and is stable.}	
\label{fig:WvsS}
\end{figure}

\begin{figure*}
\includegraphics[width=1.55\columnwidth]{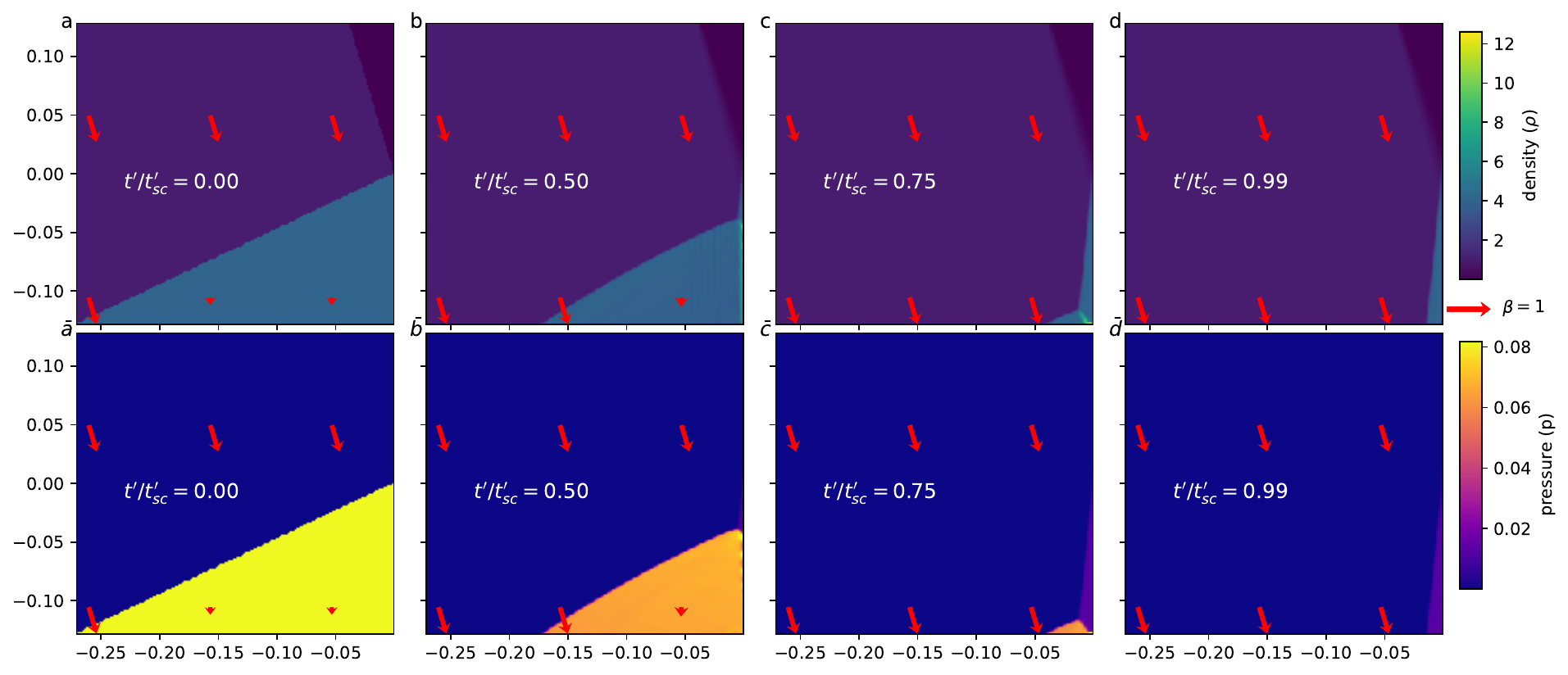}
\hspace{0.35cm}
\includegraphics[width=.45\columnwidth]{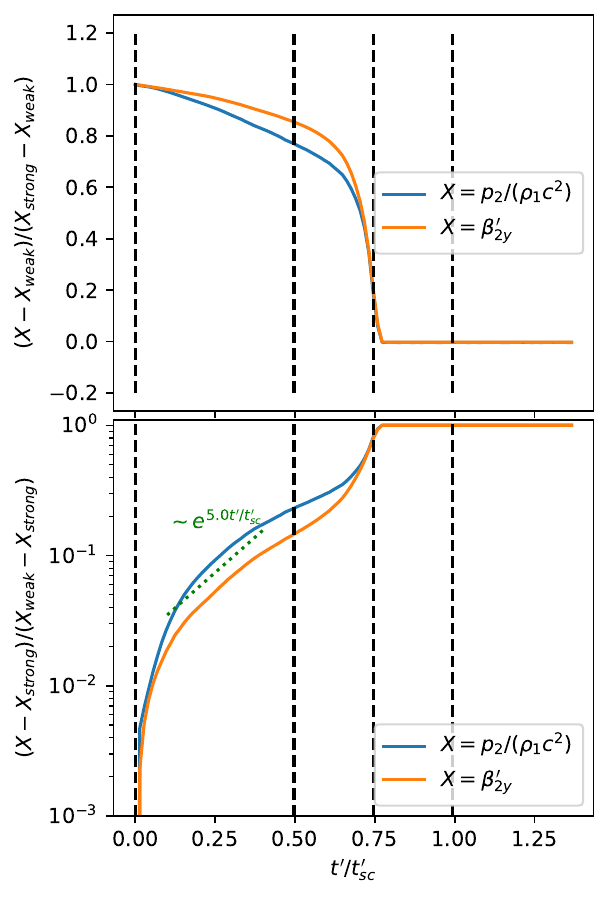}
\vspace{-0.2cm}
\caption{Stability study of the strong-shock solution, which was imposed as the initial conditions of our relativistic hydrodynamic numerical simulations performed in the steady-state frame $S'$ for the same parameters as in Figs.~\ref{fig:sim-frame-S-attached} and \ref{fig:sim-frame-Sp-attached} ($u_1=0.1$, $\alpha_1=0.306$). \textbf{\textit{Left panels}}: snapshots of the density (\textit{top panels}) and pressure (\textit{bottom panels}) at different times $t'$ in units of the sound crossing time $t'_{\rm sc}=\frac{L'}{c_{s,2}}$, where, $L'$ is the length of region~2 along the wall and $c_{s,2}$ is the region~2 sound speed in the strong-shock solution.\\ \textbf{\textit{Right panels}}: the system's spontaneous evolution from the strong-shock solution to the weak-shock solution is shown here through the fractional displacement of the normalized pressure $p_2/(\rho_1c^2)$ and vertical velocity component $\beta'_{2y}$ in region 2 from the weak-shock solution (\textit{top panel}), and from the strong-shock solution (\textit{bottom panel}).
}	
\label{fig:StoW}
\end{figure*}

\begin{figure}
\centering
\includegraphics[trim={0cm 0cm 0.0cm 0cm},clip,scale=1,width=0.48\textwidth,height=0.33\textwidth]{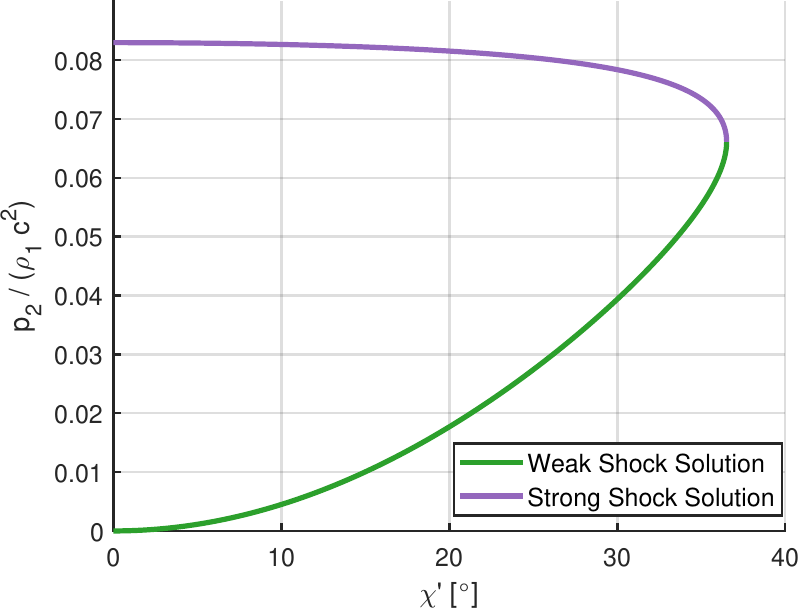}
\vspace{-0.4cm}
\caption{The shock polar for the same parameters as in Figs.~\ref{fig:sim-frame-S-attached}, \ref{fig:sim-frame-Sp-attached} and \ref{fig:StoW} ($u_1=0.1$, $\alpha_1=0.306$). For this case $\chi'=\alpha_1'=16.604^0$. For the weak (and strong) shock solution $u_2'=0.2867, \alpha_2'=17.058^0$ (and  $u_2'=0.0878, \alpha_2'=67.268^0$).}
\label{fig:shockpolar}
\end{figure}

\begin{figure}
\includegraphics[trim={0.0cm 0.85cm 1.26cm 1.82cm},clip,scale=1,width=.97\columnwidth,height=1.0\columnwidth]{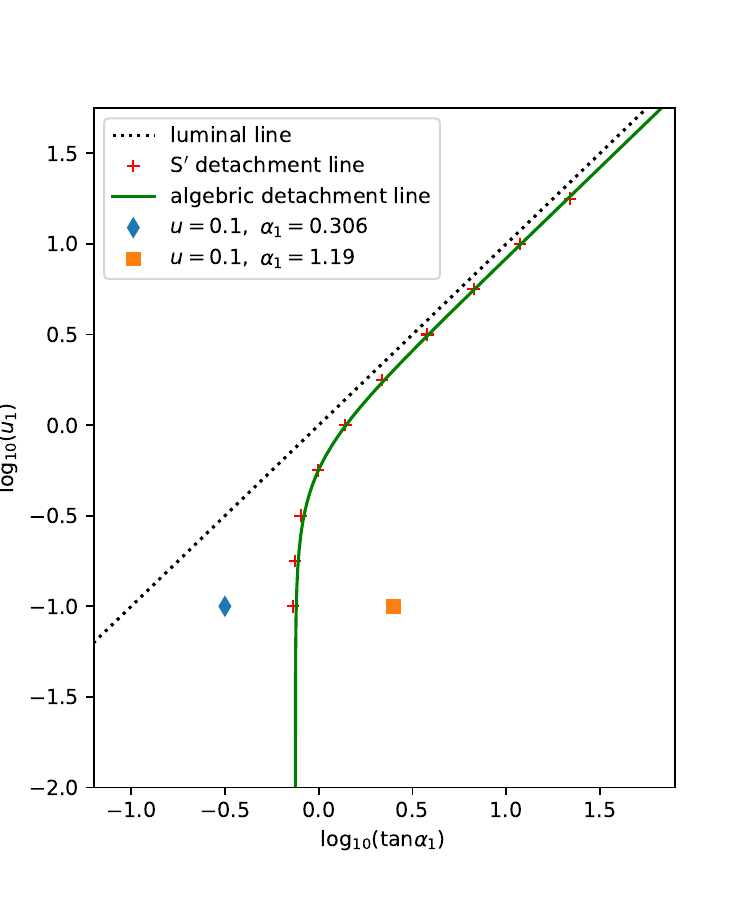}
\caption{The red `\red{+}' symbols indicate the location of the detachment line obtained from direct numerical simulation, while its analytic location is shown by the solid \textcolor[rgb]{0,0.5,0}{green} line and the luminal line is shown by the dotted black line.  For a fixed proper speed $u_1$, the interaction point $P$ remains attached to the reflecting wall for small incidence angles  ($\alpha_1<\alpha_{\rm det}(u_1)$; left of the detachment line) but detaches from the wall for large incidence angles .($\alpha_1>\alpha_{\rm det}(u_1)$; right of the detachment line). Also indicated are the parameters of the simulations corresponding to the Figs.~\ref{fig:sim-frame-S-attached}, \ref{fig:sim-frame-Sp-attached} and \ref{fig:StoW} ($u=0.1,~\alpha_1=0.306$; \textcolor[rgb]{0,0.557,0.8}{blue} diamond symbol) and Fig.~\ref{fig:sim-frame-Sp-detached} at ($u=0.1,~\alpha_1=1.19$; \textcolor[rgb]{1,0.502,0.051}{orange} square symbol).}	
\label{fig:detachment-line}
\end{figure}

Figure~\ref{fig:sim-frame-Sp-detached} shows snapshots of the density and pressure from a simulation that we have performed in the steady-state frame $S'$, for the same $S$-frame parameter values beyond the detachment line, $u_1=0.1$ but $\alpha_1=1.19$ that is larger than $\alpha_{\rm det}\approx0.649$. As expected, the interaction point $P$ detaches from the reflecting wall and shocked high-pressure material from region 2 spills into the vacuum region 0. This process also induces curvature in the shock front along with inhomogeneity in region 2.

Figure~\ref{fig:WvsS} compares the results of our relativistic hydrodynamic numerical simulations in the lab frame $S$ (blue `\blue{X}' symbols) and in the steady-state frame $S'$ (red `\red{+}' symbols) to our analytic weak-shock solution (solid \green{green} lines) and strong-shock solution (dotted \green{green} lines). The \textit{top panel} shows the compression ratio across the shock ($\rho_2/\rho_1$), the \textit{middle panel} shows the normalized pressure in region 2 ($p_2/\rho_1c^2$) and the \textit{bottom panel} shows the ratios of the tangent of the lab-frame shock-wall angle $\alpha_2$ and the incidence angle $\alpha_1$ ($\tan\alpha_2/\tan\alpha_1$). The simulations were performed for $u_1=0.1$ and several different values of the incidence angle $\alpha_1$ in the sub-luminal attachment region ($\alpha_{\rm lum}<\alpha_1<\alpha_{\rm det}$). Our initial conditions do not contain region 2, and therefore allow the system to naturally choose between the weak-shock and strong-shock solutions. In all cases, the system clearly chose the weak-shock solution, and reached a steady state implying that this solution is stable.

This motivated us to study the stability properties of the strong shock solution. Figure~\ref{fig:StoW} shows the results of a relativistic hydrodynamic numerical simulation (for $u_1=0.1$ and $\alpha_1=0.306$, the same as in Figs.~\ref{fig:sim-frame-S-attached} and \ref{fig:sim-frame-Sp-attached}) in the steady-state frame $S'$, where, for the initial conditions, we have imposed the strong-shock solution. We find that the strong-shock solution is unstable and spontaneously transitions to the weak-shock solution. After initial numerical adjustments, the volume-averaged pressure $p_2$ and vertical velocity component $\beta'_{2y}$ in the shocked region exhibit an exponentially growing deviation $\propto\exp{(5t^\prime/t^{\prime}_{sc})}$ from the strong-shock solution in the temporal evolution of the strong solution (Fig.~\ref{fig:StoW}, \textit{right bottom panel}) before approaching the weak-shock solution. This exponentially growing deviation has an $e$-folding time of about $\approx0.2t'_{\rm sc}$, i.e. one-fifth of the sound-crossing time $t^{\prime}_{sc}$ of the initial strong solution. The volume-averaged mean $p_2$ and $\beta'_{2y}$ are affected both by the gradual increase (decrease) in the size of the weak (strong) shock solution region 2, as well as by the formation of further shocks leading to denser multiply-shocked fluid between these two regions. A triple point $T$ develops at the interaction point $P$ and moves away from it at a speed of the order of the downstream sound speed ($v'_{T}\approx1.3c_{s2}$ in this simulation; see Fig.~\ref{fig:StoW}).
Above this point, the weak-shock solution is established, and as point $T$ moves away from point $P$, the weak-shock
region 2 grows, while the (higher pressure and density) strong-shock region 2 shrinks. In order to balance the pressure between these two regions, the weak-shock region 2 passes through a second shock, behind which its pressure approaches that of the strong-shock region 2, allowing for pressure equality along the CD that separates them. When point $T$ reaches the bottom edge of our simulation box and exits it only the weak-shock solution exists and steady state is established.

A qualitative explanation for the instability (stability) of the strong (weak) shock solution follows the arguments of\cite{hornung1997technical} for Newtonian shock reflection. These two solutions exist in the sub-limunal attachment region where the steady-state frame $S'$ exists, in which it is convenient to use a shock polar diagram of the downstream pressure ($p_2$) versus the deflection angle $\chi'$ (Fig. \ref{fig:shockpolar}). In terms of $p_2$ the strong (weak) shock solution is above (below) the tip of the shock polar corresponding to $\alpha'_{\rm det}=\chi'_{\rm max}$, such that it has a negative (positive) $dp_2/d\chi'=(d\chi'/dp_2)^{-1}$ and $'d\chi'/d\phi=(d\phi'/d\chi)^{-1}$, since $dp_2/d\phi'>0$ as $p_2=\frac{4}{3}\rho_1 c^2 u_{21}^2\propto u_{12}^2 = \frac{1}{4}[u_{s,1}^2-2+(4+5u_{s,1}^2+u_{s,1}^4)^{1/2}]$ naturally increases with $u'_{s,1}=u'_1\sin\phi'\propto\sin\phi'$. Now consider a small positive perturbation in $p_2$ for the strong (weak) shock solution. The slight pressure increase would cause a small increase in the shock-wall angle $\alpha'_2$ and in the shock angle $\phi'=\alpha'_1+\alpha'_2$, leading to a decrease (increase) in the deflection angle $\chi'$. This produces a downstream velocity component towards (away from) the wall, and as the wall is causally connected to the shock it increases (decreases) the pressure behind the shock $p_2$, thereby increasing (decreasing) the initial perturbation, which indicates instability (stability).
  
Figure~\ref{fig:detachment-line} compares our analytic derivation for the location of the detachment line in the $\log_{10}(u_1)-\log_{10}(\tan\alpha_1)$ plane, to the results of our hydrodynamic numerical simulations. In the latter, for each fixed value of $\log_{10}(u_1)=-1,\,-0.75,\,-0.5,\,...,\,1.25$ we have iteratively performed a series of simulations to numerically find the value of the detachment angle, $\alpha_{\rm det}(u_1)$. The agreement with our analytically derived location of the detachment line is very good.

\begin{figure*}[p]
\includegraphics
[trim={1.54cm 0.96cm 0.14cm 0.19cm},clip,scale=1,width=0.48\textwidth,height=0.56\textwidth]{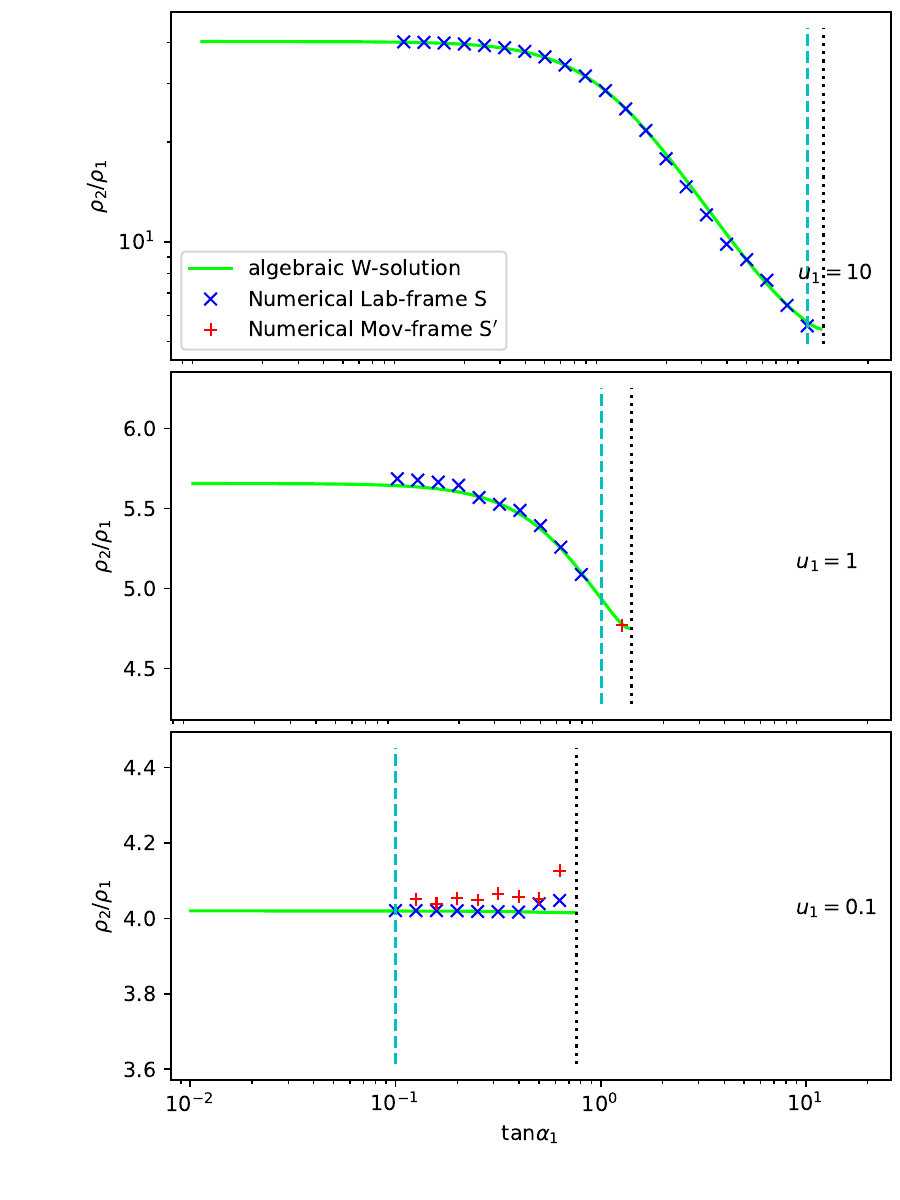}
\hspace{0.47cm}
\includegraphics[trim={0.05cm 1.45cm 1.52cm 2.43cm},clip,scale=1,width=0.48\textwidth,height=0.56\textwidth]{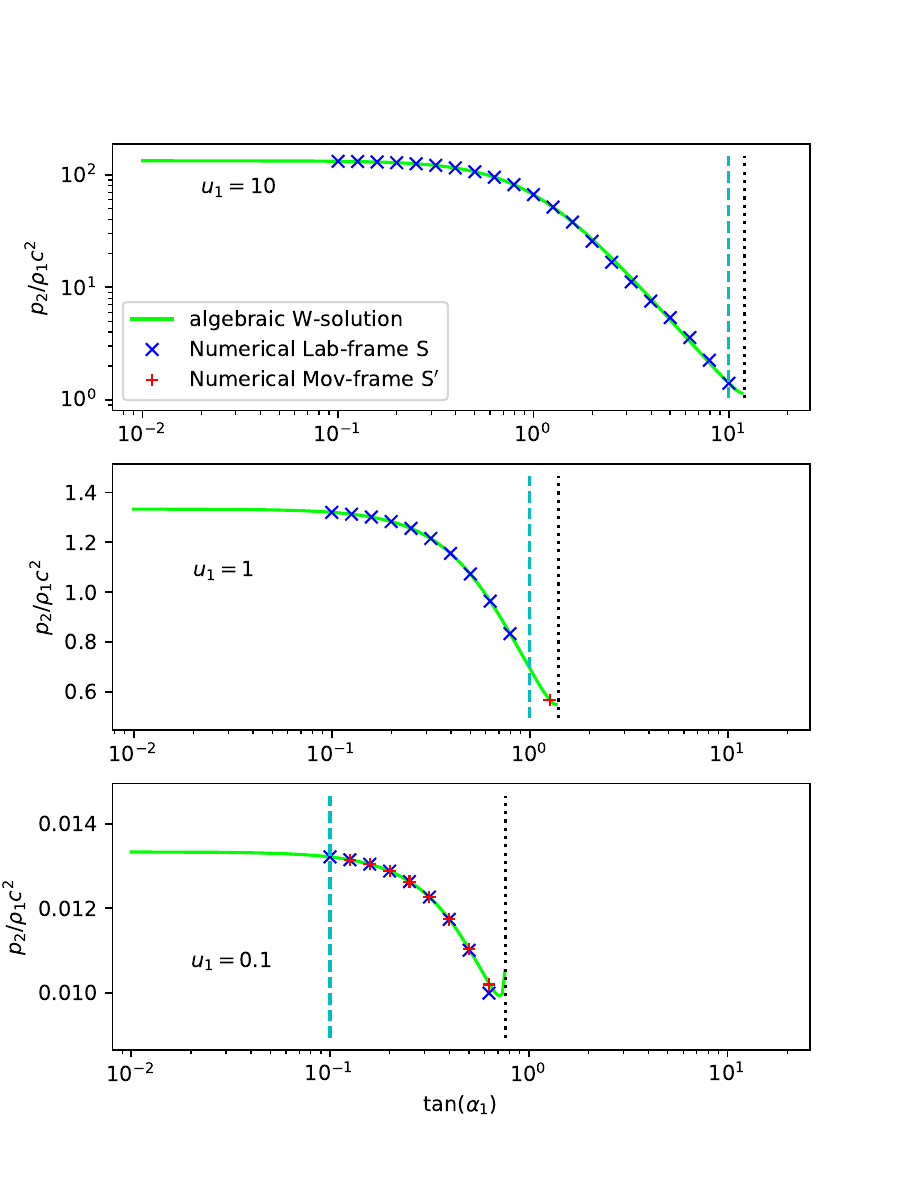}
\\
\vspace{0.8cm}
\includegraphics[trim={0.40cm 1.45cm 1.52cm 2.43cm},clip,scale=1,width=0.48\textwidth,height=0.56\textwidth]{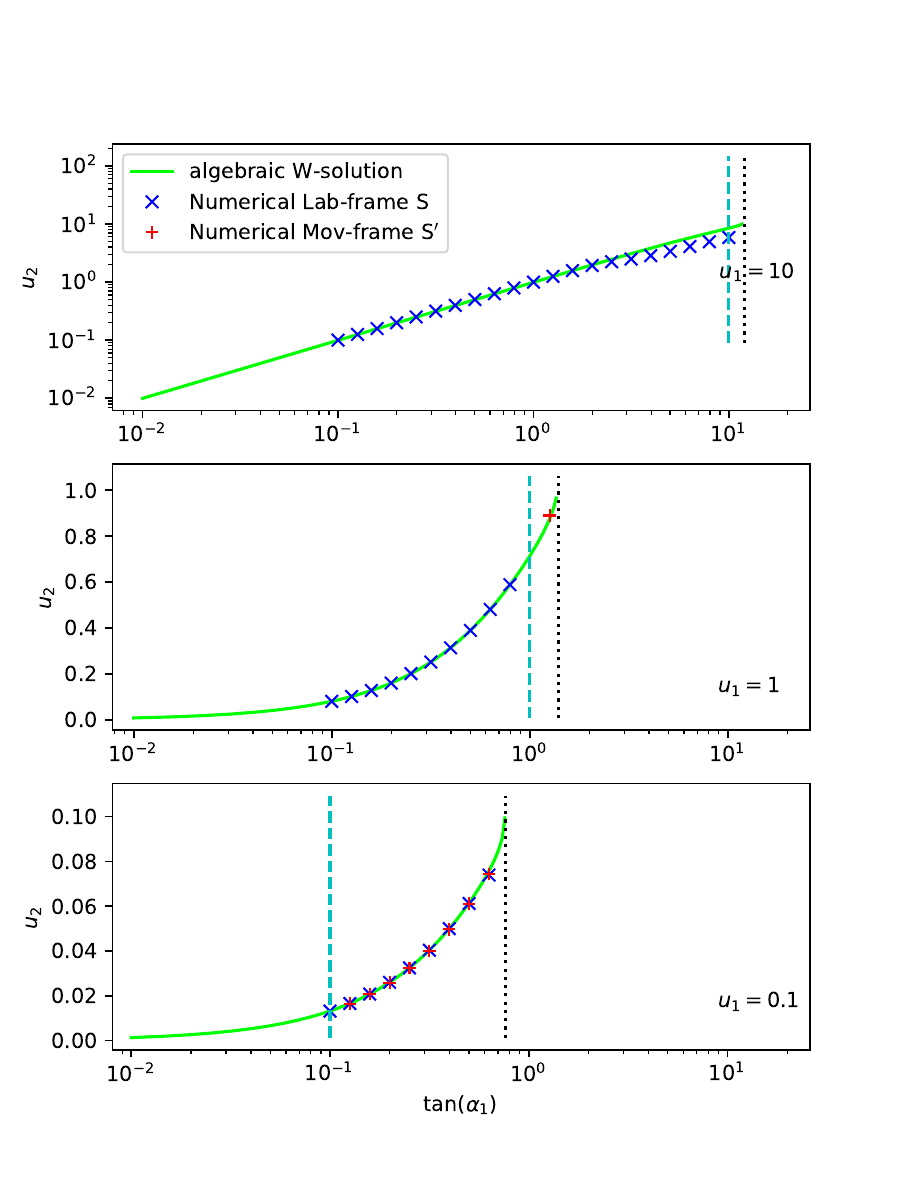}
\hspace{0.47cm}
\includegraphics[trim={1.17cm 0.95cm 0.14cm 0.19cm},clip,scale=1,width=0.48\textwidth,height=0.56\textwidth]{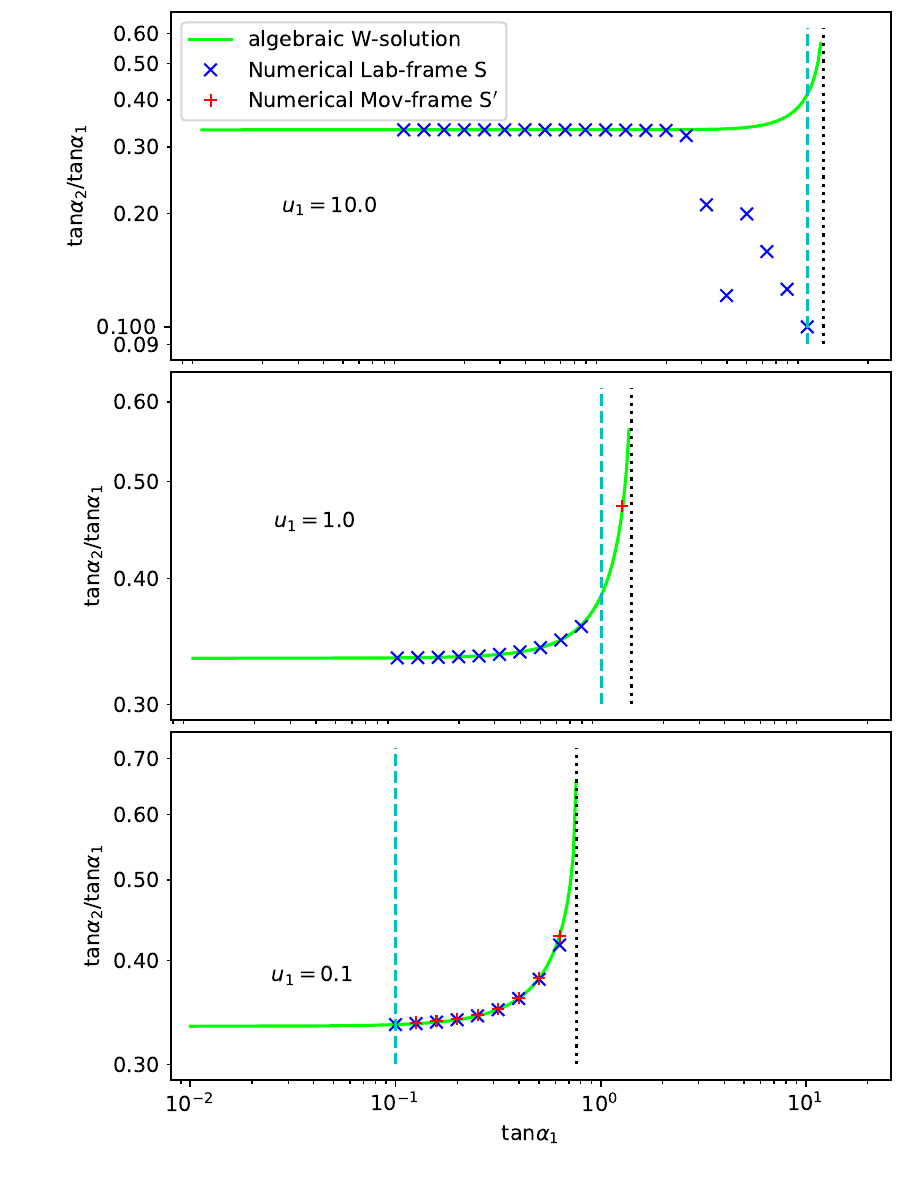}
\caption{Comparison our analytic results for the weak-shock solution (\green{green} solid lines) and our numerical simulations results in the lab frame $S$ (blue `\blue{X}' symbols) and in the steady-state frame S' (red `\red{+}' symbols) in the the sub-luminal attachment region that extends from the
luminal line at $\alpha_{\rm lum}$ (vertical dashed \cyan{cyan} line) to the detachment line at $\alpha_{\rm det}$ 
(vertical dotted black line). We show the normalized downstream density $\rho_2/\rho_1$ (\textit{top left panels}), pressure $p_2/(\rho_1c^2)$ (\textit{top right panels}), $u_2$ (\textit{bottom left panels}) and $\tan\alpha_2$ (\textit{bottom right panels}) for $u_1=0.1$ (\textit{lower panels}), $u_1=1$ (\textit{middle panels}) and $u_1=10$ (\textit{top panels}).}
\label{fig:numerical_results}
\end{figure*}

Finally, Figure~\ref{fig:numerical_results} shows a detailed comparison between our analytic results for the weak-shock solution (solid \green{green} lines) and our numerical simulation results
in the lab-frame $S$ (blue `\blue{X}' symbols) and in the steady-state frame $S'$ (red `\red{+}' symbols).
The agreement is generally very good. 
At particularly large values of $\tan\alpha_1$, corresponding to $\alpha_1\approx\pi/2$ i.e. a very grazing incidence collision, it becomes hard to accurately measure the lab-frame shock-wall angle $\alpha_2$. In our simulation of the lab-frame $S$, we have our fixed boundary conditions at the edges of the computational domain. Due to the collision of the shell with the wall a shocked region is formed near the wall, which also reaches the right part of the lower edge of our computational box, where it interferes with the boundary conditions there, causing nonphysical perturbations that propagate from that boundary into the computational box. For a large incident angle with a relativistic shell velocity, the shocked region is significantly affected by the numerical boundary condition, which significantly affects the measured values of $\alpha_2$. On the other hand, simulations in the steady-state frame $S^\prime$ remain unaffected by the boundary edge interference due to their construction (Fig.~\ref{fig:shell-wall}). Therefore, simulations in $S^\prime$ are particularly useful in the sub-luminal attachment region where both frame $S'$ and regular solutions exist.

\section{discussion}
\label{sec:dis}

This work addresses the oblique collision of a relativistic cold uniform shell with an ideal reflecting wall. We have found a fully analytic solution and study its properties both analytically and numerically. 

Similar to our recent work on relativistic shock reflection\citep{Granot-Rabinovich-24,Bera+24} we identify a super-luminal regime where no steady-state frame $S'$ exists. We adopt a similar strategy of using the integral conservation laws in the lab frame (where the cold shell's velocity is normal to its vacuum interface) along with analytic result for a 1D shock propagating into a cold medium in order to derive a fully analytic solution.

We have studied the parameter space of the cold shell's proper speed $u_1$ and incidence angle $\alpha_1$ in terms of the regular solutions where the interaction point $P$ is attached to the wall.
Similarly to our recent findings for relativistic shock reflection\citep{Granot-Rabinovich-24,Bera+24} we find three distinct regions separated by two critical lines (see Fig.~\ref{fig:critical_lines}): 
\begin{enumerate}[leftmargin=0.4cm,labelwidth=0.4cm,itemsep=-0.0em]
\item \textbf{Super-luminal} region: \textbf{one} regular solution -- weak shock,
\item \textbf{Sub-luminal attachment} region: \textbf{two} regular solutions -- weak shock \& strong shock,
\item \textbf{Detachment} region: \textbf{no} regular solutions.
\end{enumerate}
A 1D planar collision corresponds to $\alpha_1\to0$ and is therefore always in the super-luminal region where only the weak-shock solution exists, in which despite its name the shock itself can still be very strong in terms of its Mach number, i.e. $\mathcal{M}\gg1$ and in our case (for a cold upstream medium) $\mathcal{M}\to\infty$.

The weak shock solution is shown in Figs~\ref{fig:results_sin}-\ref{fig:results_norm} and strong shock solution is shown in Fig.~\ref{fig:results_tan_strong}.
The strong shock solution cannot cross the luminal line as it diverges (in terms of $\rho_2$, $p_2$, $u_2$) when it approaches the luminal line (see Figs~\ref{fig:SW_solutions_a1} and \ref{fig:SW_solutions_Gp}). 

Since the shock upstream is cold, the detachment line coincides with the sonic line (see Fig.~\ref{fig:results_6panel} and \S\,\ref{sec:cons-steady}). Along this line the weak and strong shock solutions coincide and we study the properties of the solution there in detail (see \S\,\ref{sec:sonic} and Fig.~\ref{fig:sonic-line}). We find exact analytic expressions in the Newtonian limit ($u_1\ll1$) and in the ultra-relativistic limit ($u_1\gg1$).

In the sub-luminal attachment region we study the flow in the steady-state frame $S'$ (in \S\,\ref{sec:cons-steady}). We find solutions for the shocked fluid speed $\beta'_2$ and the detachment angle that is also the maximal deflection angle, $\alpha'_{\rm det}=\chi'_{\rm max}$ (see Figs.~\ref{fig:deflection}, \ref{fig:max_deflection}). In \S\,\ref{sec:cons-steady} we also analytically prove that the detachment line and the sonic line exactly coincide for a cold upstream.

We perform relativistic hydrodynamic numerical simulations (in \S\,\ref{sec:simulations}) both in the lab frame $S$ (e.g. Fig.~\ref{fig:sim-frame-S-attached}), and in the sub-luminal regime also in the steady-state frame $S'$ (e.g. Fig.~\ref{fig:sim-frame-Sp-attached}). We find that in the sub-luminal attachment region the system naturally chooses the weak shock solution (see Fig.~\ref{fig:WvsS}), which is stable. 
In the detachment region region the interaction point $P$ detaches from the reflecting wall and hot shocked material spills into the vacuum region (see Fig.~\ref{fig:sim-frame-Sp-detached}). This is analogous to a meteorite hitting the surface of a solid planet, causing melted rock to splash away from the impact location, possibly forming chondrules in the process.
The strong shock solution is found to be unstable, and spontaneously transitions to the weak shock solution (see Fig.~\ref{fig:StoW}).
We find very good agreement between our simulation results and our analytic weak shock solution in terms of the detachment line location (Fig.~\ref{fig:detachment-line}), the flow properties in the shocked region 2 ($\rho_2$, $p_2$, $u_2$) and the shock-wall angle $\alpha_2$ (Fig.~\ref{fig:numerical_results}).

The problem solved in this work is well motivated by astrophysical objects in which relativistic outflows can lead to oblique collisions between two cold shells. The latter forms a pair of shocks each propagating into the respective cold shell where the shocked parts of the shells are separated by a contact discontinuity (CD), which may be regarded as a reflecting wall. 
Our analytic solution can be used to model the flow on each side of the CD separately, where the two sides are matched at the CD requiring equality of the pressure and normal velocity component at the CD. This reduces the problem of an oblique collision between two cold shells to two coupled problems of a single shell obliquely colliding with a wall. 

\begin{acknowledgments}
J. Granot thanks the Yukawa Institute for Theoretical Physics at Kyoto University, where part of this work was done, for their hospitality.
P. Bera thanks IoE for the grant No. R/Dev/D/IoE/Seed Grant-V/2024-25/81690.
P. Beniamini's work was supported by a grant (no. 2024788) from the United States-Israel Binational Science Foundation (BSF), Jerusalem, Israel and by a grant (no. 1649/23) from the Israel Science Foundation.
\end{acknowledgments}

\bigskip
\section*{Data Availability Statement}
The data that supports the findings of this study are available within the article.

\bibliography{bibli4}

\end{document}